\documentclass[aps,prl,twocolumn,superscriptaddress,nofootinbib]{revtex4-1}
\usepackage{amsmath}
\usepackage{graphicx}
\usepackage{amsbsy}
\usepackage{amssymb}
\usepackage{palatino,avant,graphicx,color}
\usepackage[none]{hyphenat}
\usepackage{tikz}
\usepackage{xy}
\usepackage{comment}
\usepackage[none]{changes}
\usepackage{wrapfig}
\usepackage{float}
\usepackage{tabularx}
\usepackage{natbib}
\usepackage{hyperref}

\newcommand{\cmmnt}[1]{}

\begin{document}

\title{Induced Charge Capacitive Deionization: The electrokinetic response of a porous particle to an external electric field}

\author{S. Rubin}
%\email[Email: ]{rubin.shim@gmail.com}
\affiliation{Faculty of Mechanical Engineering, Technion - Israel Institute of Technology, Haifa, Israel}
\author{M. E. Suss}
\affiliation{Faculty of Mechanical Engineering, Technion - Israel Institute of Technology, Haifa, Israel}
\author{P. M. Biesheuvel}
\affiliation{Wetsus, European Centre of Excellence for Sustainable Water Technology, Leeuwarden, The Netherlands}
% \affiliation{Laboratory of Physical Chemistry and Soft Matter, Wageningen University, Dreijenplein 6, 6703 HB Wageningen, The Netherlands}
\author{M. Bercovici}
\email[Email: ]{mberco@technion.ac.il}
\affiliation{Faculty of Mechanical Engineering, Technion - Israel Institute of Technology, Haifa, Israel}

%\today

\begin{abstract}

We demonstrate the phenomenon of induced-charge capacitive deionization (ICCDI) that occurs around a porous and conducting particle immersed in an electrolyte, under the action of an external electric field. The external electric field induces an electric dipole in the porous particle, leading to its capacitive charging by both cations and anions at opposite poles. This regime is characterized by a long charging time which results in significant changes in salt concentration in the electrically neutral bulk, on the scale of the particle. We qualitatively demonstrate the effect of advection on the spatio-temporal concentration field which, through diffusiophoresis, may introduce corrections to the electrophoretic mobility of such particles.

\bigskip
\hspace{-11pt}
% PACS numbers: 47.57.jd, 82.45.Gj, 82.47.Uv, 84.32.Tt.
DOI: 10.1103/PhysRevLett.117.234502
%PACS numbers: 04.50.Kd, 98.80.Es
\end{abstract}

\maketitle

\textbf{Introduction:} The study of electrokinetic effects dates back to the 19th century \cite{reuss1809notice, helmholtz1879studien}, and encompasses the interaction between ions, fluid flows, electrical fields, and suspended particles. 
In the past two decades electrokinetics attracted much interest in the context of microfluidic systems, due to favorable scaling of mass transport with miniaturization, which have led to a wide range of applications in bioanalysis and flow control and also stimulated theoretical investigation of novel physical regimes. 
The formation of an electric double layer (EDL) at the solid-fluid interface has been a central object of research for more than a century, and yet many aspects of its rich multiscale physics remain to be explored. Surface charge on a solid can be established by its chemical interaction with the liquid, or can be induced by an external electric field (see \cite{bazant2004induced} and references therein).
While the electrokinetic response of a polarizable impermeable particle subject to an external electric field (i.e., the induced charge mechanism) has been thoroughly investigated both theoretically and experimentally \cite{bazant2004induced, schnitzer2012induced, davidson2014chaotic, peng2014induced}, to the best of our knowledge the response of a porous polarizable particle has not been addressed to date.
%%%%%%%%%%%%%%%%%%%       Fig.1      %%%%%%%%%%%%%%%%%
\begin{figure}[ht!] %H %ht
\centering
\includegraphics[scale=0.087]{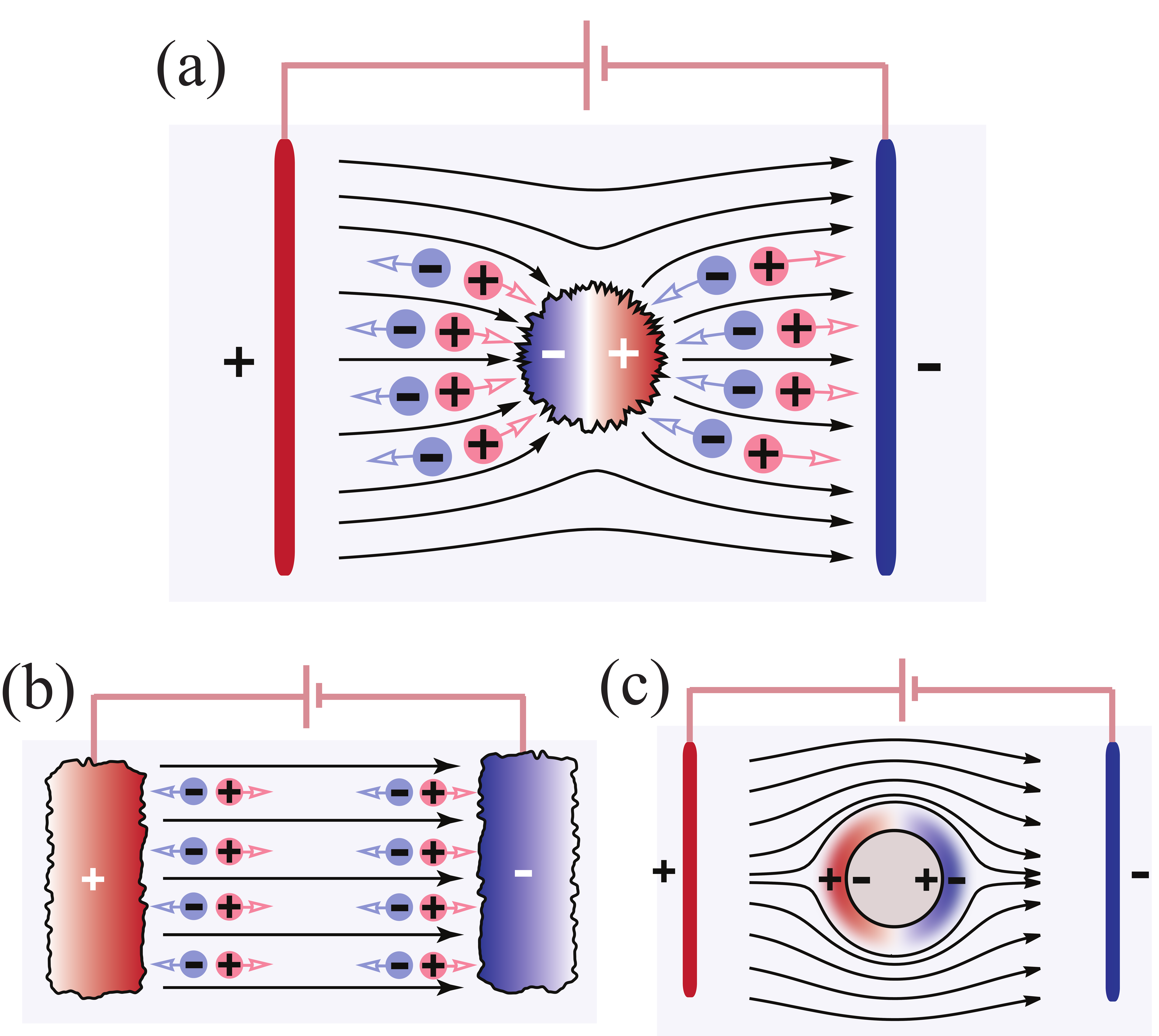}
          \caption{(a) In ICCDI, an external electric field induces an electric dipole on a porous solid which leads to its capacitive charging over long durations of time. (b) In a typical CDI configuration an electrolyte is deionized by a pair of porous conducting electrodes connected directly to a power source,  (c) in contrast to ICCDI the EDL of an impermeable and polarizable particle rapidly charges, resulting in non-penetrating electric field lines.}
\label{ConceptFigure}
\end{figure}

In this work we study the response of a conducting porous particle, characterized by a large surface to volume ratio, to an externally applied DC electric field. Owing to the large surface area, the polarization of the particle's surface leads to a new physical regime in induced charge electrokinetics, characterized by a long charging time and non-linear dynamics in the electrically-neutral bulk, which generates large depletion regions (on the order of the particle). The charging process can be described by source terms in the porous particle, and when coupled with electromigration, diffusion and advection (e.g. pressure driven, or by induced charge electroosmosis) results in spatial distributions of salts, which are fundamentally different from those obtained in capacitive charging of impermeable particles. We experimentally investigate this regime around a fixed particle using a binary electrolyte in which one of the ions is fluorescent, and provide a two-dimensional model which qualitatively captures key properties of this process. Our analysis also indicates the importance of such electrosorption on the electrophoresis of polarizable porous particles.

The first studies on ion-transport within and around porous electrodes \cite{johnson1971desalting, de1963porous, newman1962theoretical, newman1975porous} were initiated in the 1960's. These processes, commonly  referred to as capacitive deionization (CDI), are of great interest for their potential applications \cite{suss2015water,dunn2000predictions,kotz2000principles,simon2008materials,brogioli2009extracting,rica2012thermodynamic} which include water desalination and energy storage. Fig.(\ref{ConceptFigure}) illustratively compares the induced charge capacitive deionization (ICCDI) regime, which is the subject of our study, with the cases of CDI and of an impermeable induced charge particle. In standard CDI (Fig.(\ref{ConceptFigure}b)), a power source is connected to two separate porous electrodes, such that one acts as a cathode and the other as an anode. The deionization processes occurs as negative and positive ions electromigrate towards the anode and cathode, respectively. 
Similarly, in the ICCDI regime positive and negative ions, from around and inside the porous particle, are electrosorbed (or expelled) at its two oppositely charged regions. Fig (\ref{ConceptFigure}c) presents the process of induced charge around an impermeable conducting particle at the low Dukhin number ($\text{Du}$) regime \cite{bikerman1940electrokinetic,dukhin1993non}. In this regime the EDL quickly achieves equilibrium, and changes to ionic concentrations are limited to the EDL. At high surface conductance ($\text{Du} \sim O(1)$ or higher), significant concentration polarization arises, characterized by enrichment regions perpendicular to the applied electric field and by depletion regions parallel to it \cite{leinweber2006continuous,davidson2014chaotic}, accompanied by penetration of charge into the bulk.  
In marked difference, the ICCDI regime results in continuous growth of large depletion regions (on the order of the particle's size) in the electrically-neutral bulk which are the result of the particle's large surface to volume ratio. Notably, this regime is independent of surface conductance and holds even for $\text{Du} \ll 1$ and moderate electrical fields.

\textbf{Theoretical analysis:} We begin by considering ion transport in porous media with a bimodal pore size distribution, characterized by a hierarchical structure having two types of pores. For activated carbon, relevant for our study, these are electro-neutral macropores of a typical scale of $1 \mu$m, and electrically charged micropores with overlapping EDLs of a typical scale of $1$ nm, which occupy regions of porosities $p_{M}$ and $p_{m}$, respectively \cite{johnson1971desalting, biesheuvel2012electrochemistry,eikerling2005optimized, suss2014situ,hemmatifar2015two, biesheuvel2015theory, huang2012relation}.
In our notation, the subscripts $m,M,B$ indicate a physical quantity within the distinct regions of the micropores (m), macropores (M), and bulk (B), whereas $+,-$ distinguish between cations and anions. In the macropores and the bulk, the current density of ionic species in a binary and symmetric electrolyte ($z=z^{\pm}>0$), due to diffusion, advection field $\vec{u}$ and electro-migration is given by 
\begin{equation}
	\vec{J}^{\pm}_{M,B} = -\left( D_{M,B}^{\pm} \vec{\nabla}c_{M,B}^{\pm} \mp F z b_{M,B}^{\pm} c_{M,B}^{\pm} \vec{\nabla} \varphi_{M,B}^{ }- c_{M,B}^{\pm} \vec{u} \right),
\label{Fluxes}
\end{equation}
where $F$ is the Faraday constant, $\varphi$ is the electrostatic potential, and $b_{}^{\pm}$ and $D_{}^{\pm}$ are, respectively, the effective electrophoretic mobilities (which account for finite dissociation and ionic strength effects), and the diffusion coefficients. In light of better agreement with some experimental regimes \cite{biesheuvel2012electrochemistry,suss2014situ,hemmatifar2015two,biesheuvel2015theory}, we here 
adopt the modified Donnan (mD) model for capacitive charging which assumes no transport in the micropore region,
%%%%%%%%%%%%%%%%%%%       Fig.2      %%%%%%%%%%%%%%%%%
\begin{figure}[ht] %ht
\centering
\includegraphics[scale=0.36]{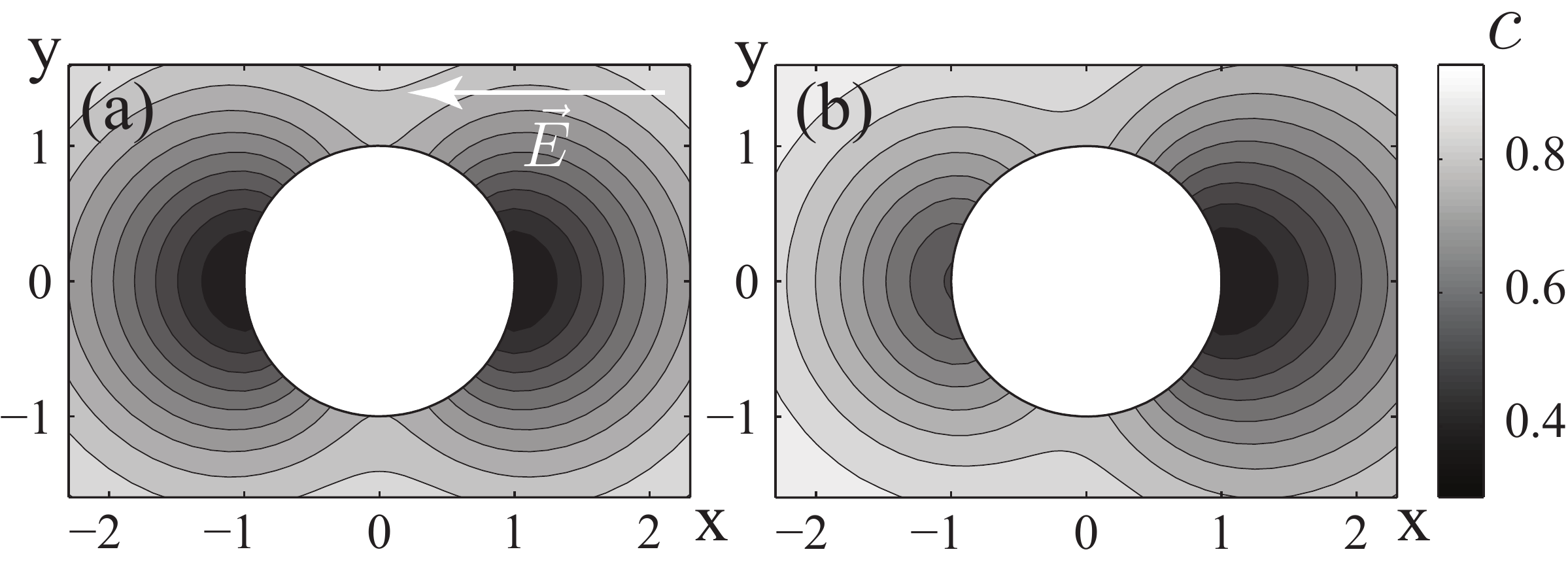}
          \caption{Numerical simulation results at a non-dimensional time $0.5$ (scaled by $a^{2}/D$) \cite{SuppInfo}, showing changes in the initial uniform concentration distribution ($c_{0}=1$) due to electrosorption invoked by ICCDI around a porous disk. (a) presents the symmetric case, while (b) shows the asymmetric case where the positive ion has a higher mobility than the negative ion, and a larger depletion region forms around the negative pole.}
\label{Numerics}
\end{figure}
but note that the qualitative results of our study remain unchanged when using the Gouy-Chapman (GC) model.
The governing Nernst-Planck equations for ionic species in the macropore region take the form 
\begin{equation}
	\frac{\partial c_{M}^{\pm}}{\partial t}+ 	\frac{1}{p_{M}^{}}\vec{\nabla} \cdot \left( p_{M} \vec{J}^{\pm}_{M}  \right) =
-I_{S},
\label{NPWithDonnan}
\end{equation}
where $I_{S}$ is a source term representing the consumption of ions in the micropores. 
%$I_{S}=-\frac{p_{m}}{p_{M}} e^{-\frac{\Delta \mu}{e V_{T}}} \cdot \partial \left( c_{M}^{} \cosh \left( \frac{ z \Delta \varphi_{d}}{V_{T}} \right) \right) / \partial t$,
\cite{biesheuvel2012electrochemistry} (see \cite{SuppInfo} for an explicit expression). A similar equation with $p_{M}=1$ and $I_{S}=0$ holds in the bulk.
The potential difference between the electrode surface and macropore space, $\varphi_{e}-\varphi_{M}$, is the sum of Donnan potential ($\Delta \varphi_{d}$) which represents the total potential drop between the macropores and the micropores, and the Stern potentials ($\Delta \varphi_{s}$) which represents the potential jump between the micropores and the surface (see \cite{SuppInfo} for an explicit expression),
\begin{equation}
	\varphi_{e}-\varphi_{M}=\left( \varphi_{e}-\varphi_{m} \right)+(\varphi_{m}-\varphi_{M})=\Delta \varphi_{s}+\Delta \varphi_{d}.
\label{DonnanRelatesToOther}
\end{equation}
%where $\Delta \varphi_{s}=\dfrac{2 F z}{C_{s}}  e^{-\frac{\Delta \mu}{F V_{T}}} c_{M} \sinh \left( \dfrac{z \Delta \varphi_{d}}{V_{T}} \right)$.
%Here, $C_{s}$ is the Stern layer capacitance and 
For convenience, we express Eq.(\ref{Fluxes},\ref{NPWithDonnan},\ref{DonnanRelatesToOther}) as a function of the (half) neutral salt concentration and (half) charge density defined by $c=(c^{+}+c^{-})/2$ and $\rho=Fz(c^{+}-c^{-})/2$. The dynamics in the bulk and macropore regions is linked through matching conditions which stem from mass and charge conservation (see discussion in \cite{SuppInfo}).

Before turning to numerical solutions of the nonlinear governing equations for ion-transport Eq.(\ref{Fluxes},\ref{NPWithDonnan},\ref{DonnanRelatesToOther}), we seek to gain some insight by analyzing the angular distribution of the salt concentration and of the electrostatic potential at early times. To this end, we consider again the case of no advection ($\vec{u}=0$), consider a symmetric case of equal diffusion coefficients, $D_{M,B}^{+}=D_{M,B}^{-}$, and
perform Taylor expansion of salt concentration and electric potential up to second order in time $\delta t$ via, $c = c_{0} + \delta c + \delta^{2}c +\ldots$, and $\varphi = \varphi_{0} + \delta \varphi + \delta^{2} \varphi+ \ldots$. The corresponding relations 
that couple  $\delta^{2} c$ and $\delta \varphi$ take the form
% \begin{minipage}{.55\linewidth}
% \begin{equation}
%   \dfrac{\partial \delta^{2} c}{\partial t} - \nabla^{2} \delta^{2}c = A \delta \varphi \dfrac{\partial \delta \varphi}{\partial t}, \vspace{0.1in} 
%   \label{LinearizedUpToSecondOrdera}
% \end{equation}
% \end{minipage}%
% \begin{minipage}{.45\linewidth}
% \begin{equation}
%   \nabla^{2} \delta \varphi = B \dfrac{\partial \delta \varphi}{\partial t}. \vspace{0.1in} 
%   \label{LinearizedUpToSecondOrder}
% \end{equation}
% \end{minipage}
\begin{subequations}
\begin{align}
	\nabla^{2} \delta \varphi &= B \dfrac{\partial \delta \varphi}{\partial t}
		\\
	\dfrac{\partial \delta^{2} c}{\partial t} - \nabla^{2} \delta^{2}c &= A \delta \varphi \dfrac{\partial \delta \varphi}{\partial t},
\end{align}
\label{LinearizedUpToSecondOrder}
\end{subequations}
and are in principle model independent. $A$ and $B$ are positive coefficients, that can be calculated for both GC and mD capacitive charging models \cite{SuppInfo}.
Note that the term $\partial \varphi / \partial t$ can be eliminated by substituting Eq.(\ref{LinearizedUpToSecondOrder}a) into Eq.(\ref{LinearizedUpToSecondOrder}b), resulting in an equation which directly relates the leading term of the potential $\delta \varphi$ to the leading term of the concentration $\delta^{2}c$, $\partial \delta^{2} c/\partial t - \nabla^{2} \delta^{2}c = A/B \cdot \delta \varphi \nabla^{2} \delta \varphi$.
At short times, $\delta \varphi$ admits an initial behavior of a dipole, i.e.  $\cos(\theta)$ dependence, where $\theta$ is the angle with respect to the horizontal axis, in a coordinate system concentric with the disk. 
This angular dependence of $\delta \varphi$ serves as a consistent initial condition for Eq.(\ref{LinearizedUpToSecondOrder}a). Since tangential derivatives in concentration along the edge of the disk are expected to be much smaller than radial ones, the tangential components in the Laplacian of 
Eq.(\ref{LinearizedUpToSecondOrder}b) can be neglected, indicating that $\delta^{2} c$ must admit an angular dependence of $\cos^{2}(\theta)$. The most significant depletion regions in the electroneutral bulk are thus anticipated at the poles ($\theta=0,\pi$) of the disk,
which also occurs for diffuse charge distribution in the EDL around an impermeable and ideally polarizable disk. \cite{chu2006nonlinear}

% \textcolor{blue}{Interestingly, at least for electroadsrorption processes which are slow relative to the ambipolar diffusion time scale, the asymmetric case experience additional $\vec{E}_{D}=V_{T}\frac{D^{+}-D^{-}}{D^{+}+D^{-}}\frac{\vec{\nabla}c}{c}$  
% we are expecting ...}
%%%%%%%%%%%%%%%%%%       Fig.3      %%%%%%%%%%%%%%%%%
% At later times, non-linear behavior leads to non-separable solutions which hampers analytical treatment. 
To obtain the dynamics over longer times we turn to a two dimensional numerical simulation. To this end, we use a finite elements software (\begin{small}{COMSOL}\end{small} Multiphysics \cite{Comsol}), to solve the set of Nernst Planck Equations Eq.(\ref{NPWithDonnan}) and the mD model for capacitive charging (see \cite{SuppInfo} for detailed information on the simulation).
% \deleted{incorporated through Eq.(\ref{DonnanRelatesToOther}})  
For simplicity, we first focus on the case of $\vec{u}=0$.
% (see \cite{SuppInfo} for detailed information on the simulation, including boundary conditions, grid definitions, initial conditions, and convergence results). 
% The value of the floating potential, $\varphi_{e}$, on the porous electrode may in principle depend on time (see \cite{SuppInfo} for discussion), and in our work we for convenience set $\varphi_{e}=0$ at all times.
% , which is expected to hold when the disk is centered between the electrodes and $a/L \ll 1$. For $a/L \sim 1$, left/right asymmetry in concentration and conductivity lead to an asymmetric potential drop and therefore a time dependent value of $\varphi_{e}$, even for a centered porous particle.
% \deleted{The initial configuration is assumed to be: uniform concentration of ionic species, dipole electric potential outside the disk and zero withing the porous region, zero Donnan potential, and zero time derivatives of $\partial c/ \partial t \vert_{t=0}$, $\partial \varphi/ \partial t \vert_{t=0}$ and $\partial \Delta \varphi_{d}/\partial t \vert_{t=0}$.} 
Fig.(\ref{Numerics}a,b) present numerical solutions of the salt concentration distribution for the cases of a symmetric and a non-symmetric electrolyte, respectively. Consistent with our analysis for early times, we indeed obtain the most significant depletion in the vicinity of the two poles.
%%%%%%%%%%%%%%%%% Fig.3 %%%%%%%%%%%%%%%%%%%%
\begin{figure}[ht] %ht! %t!
\centering
\includegraphics[scale=0.27]{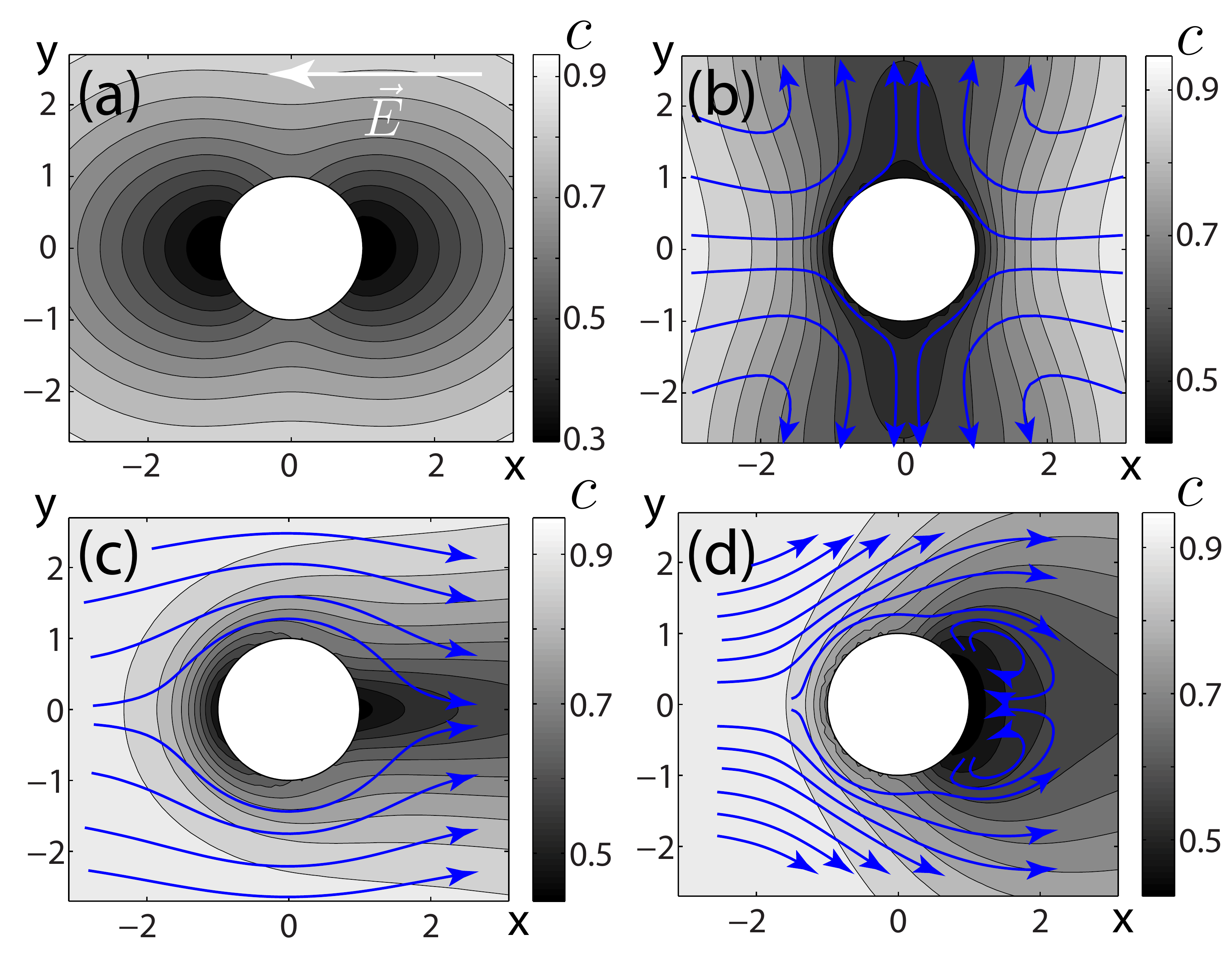}
          \caption{Numerical simulation results at a non-dimensional time $1$ (scaled by $a^{2}/D$) \cite{SuppInfo}, showing the concentration distribution due to the combined effect of ICCDI, uniform flow (either pressure driven or electroosmotic) and ICEO on a porous cylinder. 
(a) In the absence of advection, depletion regions are formed at $\theta=0,\pi$. (b) With strong ICEO, convection shifts the depletion toward the $\theta=\pm \pi/2$. (c) Under uniform flow, the left depletion region diminishes, while the right one is extended. 
In (b) and (c) the maximum slip velocity is set as $16 D/a$.
(d) Illustration of the combined effect of ICEO and advection.}
\label{AdvectionICEO}
\end{figure}
%%%%%%%%%%%%%%%%%%       Fig.4      %%%%%%%%%%%%%%%%%
\begin{figure*}[t] %ht! %t!
\centering
\includegraphics[scale=0.40]{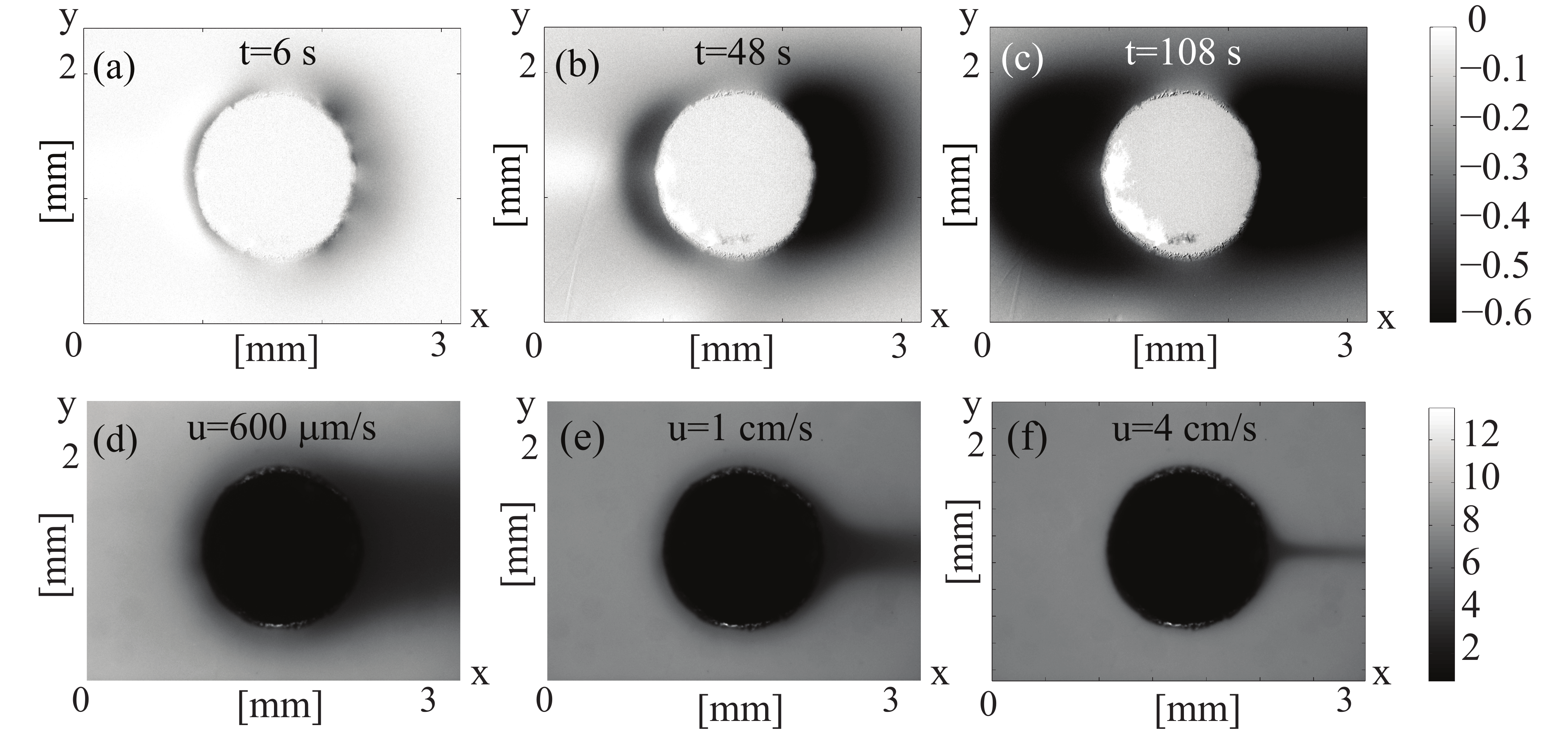}
          \caption{Experimental results showing the concentration field due to ICCDI.  Images show fluorescence signal of $100 \text{ } \mu$M sodium fluorescein under an applied potential difference of $20 \text{ } V$ between the right and left electrodes, around a $1.2 \text{ }$mm diameter carbon disk at times (a) $6 s$, (b) $ 48 s$ and (c) $108 \text{ }$s. Each frame (a-c) presents the change in fluorescence relative to the first frame, and is normalized by it for flat-field correction. (d-f)  raw fluorescence images showing the combined effect of ICCDI with pressure driven flow from left to right, at different flow velocities $u$; (d)  600 $\mu$m/s cm/s, (e) 1 cm/s, (f) 4 cm/s.}
\label{ICCDImatrix}
\end{figure*}
% and as anticipated, changes of concentrations diffuse from the surface $r=a$ where $\nabla^{2}\varphi$ acquires its largest values \cite{SuppInfo}. 
% \textcolor{blue}{Assuming that positive ion (positive) has larger effective electrophoretic mobility of than Fluoroscene ion (negative), the larger depletion region is around the negative pole in both cases.}
Furthermore, for the asymmetric case, characterized by the positive ions having a higher effective electrophoretic mobility than the negative ions, the larger depletion region, presented in Fig.(\ref{Numerics}b), forms around the negative pole.

For an impermeable polarizable particle, the charging time, $\tau_{ch}$, is on the order of the charge relaxation time $\lambda_{D}a/D$ 
(where $\lambda_{D}$ is the Debye length scale \cite{zangwill2013modern}), and introduces concentration changes on the order of $\sqrt{\lambda_{D}/a}$ over a narrow region of $\sqrt{\lambda_{D}a}$ \cite{bazant2004diffuse}. 
For $a=1$ mm, $\lambda_{D}=100$ nm, and $D=10^{-9}$ m${}^{2}$/s this corresponds to $\tau_{ch}= 100$ ms and a $1 \%$ change in concentration over a $10$ $\mu$m region. 
% The Dukhin number in this case is equivalent to the ratio $\tau_{RC}/ \tau_{D}$ \cite{bazant2004diffuse}, and typically $\ll 1$.
In contrast, while the charge relaxation time in ICCDI remains unchanged, 
% \deleted{(governed by the scale of the particle)} \textcolor{blue}{($\lambda_{D}a/D$)},
%%%%%%%%%%%%%%%%%%%%%%%%%%%%%%%%%%%%%
%\deleted{after a short transition time, $\tau_{TL}$, depletion of the bulk begins to limit the availability of the salt.
%The time scale for $\tau_{TL}$ can be obtained from scaling of the governing Eq.(\ref{NPWithDonnan}), (see \cite{SuppInfo} for derivation) and is given by $\tau_{TL}=p_{m}/p_{M} \cdot a^{2}/D \cdot e^{-\Delta \mu/F V_{T}}$,
%which is analogous to the transmission line time scale obtained from the GC model, $\lambda_{D}/h_{p} \cdot  l_{p}^{2}/D$,
% \cite{biesheuvel2010nonlinear, mirzadeh2014enhanced}.
%For $p_{m}/p_{M} \sim 1$, $\Delta \mu=0$, $a=1$ mm and $D=10^{-9}$ m$^{2}$/sec, $\tau_{TL}=1000$ sec.}
%%%%%%%%%%%%%%%%%%%%%%%%%%%%%%%%%%%%%%%%
% For $\lambda_{D}=10$ nm, $h_{p}=50$ nm, $l_{p}=1$ mm and $D=10^{-9}$ m$^{2}$/sec the latter is given by $\tau_{TL}=200$ sec.
the actual charging time, $\tau_{ch}$ 
depends both on the availability of ions around the porous particle
and on their propagation within the porous region. %The latter can be estimated from the electric field equation for short times (see \cite{SuppInfo}), and is given by the transmission-line time scale $\tau_{TL}=p_{m}/p_{M} \cdot a^{2}/D \cdot e^{-\Delta \mu/F V_{T}}$ \cite{biesheuvel2010nonlinear, mirzadeh2014enhanced}.  
However, for a diffusion limited process $\tau_{ch}$ is simply determined by the diffusion time scale in the bulk, $\tau_{D}=a^{2}/D$.
In cases where advection is present the availability of ions increases and charging rate grows. 
Fig.(S5) in \cite{SuppInfo} shows the flux of salt into the porous disk as a function of time,
at different slip velocities, for both dipolar and quadrupolar flows.
% values of Peclet number, defined as $\text{Pe}=a U/D$}, \underline{where $U$ is the velocity magnitude of the incoming flow (see} \underline{Fig.(\ref{Numerics}c)), far from the particle.
%Our simulation results (see \cite{SuppInfo}) further support that a charging process evolves on a typical time scale $\tau_{D}$, with electric field lines having a significant component perpendicular to the particle's outer surface at early times ($t \ll \tau_{D}$), which reduces at later times as the charging process proceeds. 

%\textcolor{blue}{The difference between an impermeable polarizable particle and a porous one is further evident when considering the effects of convection, such as pressure driven flow or electroosmosis (fixed or induced).  
%For an impermeable particle, the charging time is on the order of the charge relaxation time.
%Since the charging dynamics of the porous particle result in a non-uniform salt distribution which evolves over a large period of time, 
Beyond aspects for charging time, advection in the bulk affects the spatial distribution of salt. This is an inherently unsteady process, in which the time-dependent electrosorption  operates at similar rates as diffusion and advection.  
Fig.(\ref{AdvectionICEO}) presents 
numerical simulation results 
showing the concentration distribution for 
ICCDI cases which also include advection.
We show the cases of dipole and quadrupole flows, which correspond, respectively, to the cases
of native electro-osmosis and induced charge electro-osmosis (ICEO) \cite{bazant2004induced}.

It is worth noting the relevance of ICCDI to electrophoresis and diffusiophoresis of mobile porous (and polarizable) particles. In particular, the self-generated salt gradient over the scale of the particle introduces a retardation force due to osmotic pressure in the direction of $\vec{\nabla} c$ (the so-called chemiophoretic term \cite{dukhin1976electrophoresis,prieve1984motion}), 
and an additional electric force 
which stems from the difference in diffusivities of the cations and the anions. The latter generates an additional electrophoretic term given by
$\frac{3}{2 z^{2}} \frac{D^{+}-D^{-}}{D^{+}+D^{-}} \frac{k_{B}T}{6 \pi \eta \lambda_{B}} \frac{\zeta}{V_{T}} \vec{\nabla} \log(\frac{c}{c_{0}})$ 
which, depending on the sign of $\zeta \cdot (D^{+}-D^{-})$, is directed with or against the direction of
$\vec{\nabla} c $ \cite{prieve1984motion}.
Typical values of $\zeta=75 \text{ } mV$, $\frac{k_{B}T}{6 \pi \eta \lambda_{B}}=350 \text{ }\mu m^{2}/s$, $\lambda_{B}=0.7 \text{ }nm$, and $\Delta c/c=1/10$ lead to velocities on the order of $100 \text{ } \mu m/s$ for a $100 \text{ }\mu m$ diameter particle. 
% Notably, while the electrophoretic mobility of an impermeable polarizable particle is retarded by concentration polarization which emerges at sufficiently high surface conduction,  the asymmetric salt gradient of a porous particle is fundamentally different and may result in either retardation or advancement. 
Notably, relatively high electric fields are required in order to activate the non-linearity which, through surface conduction, results in retardation of an impermeable polarizable particle \cite{dukhin1993non}. In contrast, the asymmetric salt gradient around a porous particle gives rise to non-linearities even at low fields, and may result in either retardation or advancement.

\textbf{Experimental results:} we experimentally investigated the process of ICCDI by placing a disk-shaped activated porous electrode ($1$ mm diameter) in an acrylic microfluidic chamber (W$\times$L$\times$H = $15$ mm $\times$ $75$ mm $\times$ $250$ $\mu$m) containing a binary electrolyte solution of $100$ $\mu$M sodium fluorescein. 
We applied an external electric potential difference of $20 \nobreak \text{ V}$ from electrodes situated in two reservoirs located at the far ends of the chamber $75$ mm apart (the estimated value of a uniform electric field component, $E_{0}$, is $260 \nobreak \text{ V/m}$). 
% The electric field is sufficiently low to avoid Faradaic reactions on the surface of the disk (such reactions are clearly observable through the production of gas bubbles at higher electric fields). 
This setup is mounted on top of an inverted epifluorescence microscope (see \cite{SuppInfo} for complete details of the setup), where we image the fluorescence intensity at time intervals of $600$ ms over a total duration of $15$ min. Since at these concentrations the fluorescence intensity is proportional to concentration, it provides an indication for the concentration of the negative ion. 

Fig.(\ref{ICCDImatrix}a-c) present the fluorescence intensity around a single porous particle at different times in the charging process. At short times ($t=6$ s), a thin depletion region is formed around the disk. Notably, depletion is more significant around the poles of the disk, as predicted by the short-times analysis. At later times, ($t = 48 \nobreak \text{ s}$), the asymmetry in the size of the depletion regions is clearly visible, as expected from the difference in mobility between sodium and fluorescein, and as predicted by the numerical simulation. We note that another source of asymmetry is the electro-osmotic flow on the chamber's walls which is directed along the electric field lines, and acts to extend the depletion region around the negative pole. After $108$ s, the depletion region is on the scale of the disk, and after another $\sim 10 \nobreak \text{ min}$ the ionic flux from the surrounding bulk is balanced with the charging rate of the micropores, which leads to a quasi-steady regime characterized by nearly static depletion regions (see \cite{SuppInfo} for data set). Consistent with our estimations for $\tau_D$, we indeed observe in our experiments (with no advection) charging times of tens of minutes and much more significant changes in concentration ($\sim 1/3$) over larger distances ($\sim a$) as compared to impermeable particles.   
In the Supplementary Information \cite{SuppInfo} we present a similar time-lapse experiment performed on a staggered array of disks (Fig S6), and the discharge of fluorescein when the electric field is flipped (Fig S8). 
In a presence of advection (Fig.(\ref{ICCDImatrix}d), flow velocity $600$ $\mu$m/s), the charging time reduces to approximately $15$ min, as indicated by gradual disappearance of the the downstream deletion wake.
%showing that a relatively large volume of the bulk can be processed using a set of porous electrodes.
%Fig.(\ref{ICCDImatrix}d-f) presents a similar time-lapse experiment performed on an array of $18$ porous disks arranged in a staggered array, with a typical distance of $1$ mm between the disks. At $78$ s, clear interaction between the depletion fronts of the individual disks is observed, and by $240 \nobreak \text{ s}$ a continuous depletion region exists between the disks. i.e. a relatively large volume of the bulk can be processed using a set of porous electrodes, each sufficiently small to operate in an (induced) capacitive mode. %Reversing the direction of the electric field leads to discharge of the Fluorescein, and increased fluorescence around one of the poles (see \cite{SuppInfo} for data set).

\textbf{Summary, conclusions and future directions:} 
We studied the electrokinetic response of a conducting porous particle to an externally applied electric field. 
As demonstrated both by our experimental and numerical results, the ICCDI phenomenon is characterized by charging time which is several orders of magnitude larger than that of a polarizable impermeable particle, and leads to significant changes in salt concentration in the electrically-neutral bulk. Consequently, in ICCDI the processes of electrosorption, electromigration, diffusion and advection, are strongly coupled as they operate on similar time scales.

Several non-linear effects are triggered by the strong electrosorption of the porous particle, which merit further investigation. In the advection-free case, we observed sharp concentration fronts propagating away from the particle which are likely associated with conductivity and pH gradients induced by the particle. The formation of these gradients is particularly interesting as those affect the electrophoretic mobility of the participating ionic species and thus couple back to electromigration and electrosorption fluxes. Modeling of such multi-coupled processes requires construction of more elaborate numerical schemes.

The effects we observed in this work may be particularly important when considering the electrophoretic mobility of such particles. Most importantly,  the self-generated salt concentration gradient around the particle is expected to result in significant diffusiophoretic forces which, depending on the species' diffusivities and the zeta potential of the surface, may either retard or advance the particle. Furthermore, the above-mentioned pH changes may also have a significant influence on the native zeta potential of the surface and also affect its mobility.

From a practical perspective, ICCDI may be useful for the implementation of novel desalination methods, as it allows rapid removal of ionic species without requiring physical connection of the electrode. 

\textbf{Acknowledgments:} S.R. is supported in part by a Technion fellowship from the Lady Davis Foundation. We gratefully acknowledge the assistance of Mr. Eric Guyes in building the fluidic cell, and thank Prof. Martin Bazant for useful comments. 

%%%%%%%%%%%%%%%%%%%%%%%%%%%%%%%%%%%%%

%

%%%%%%%%%% Merge with supplemental materials %%%%%%%%%%
\pagebreak
\widetext
\begin{center}
\textbf{SUPPLEMENTAL INFORMATION \\
Induced Charge Capacitive Deionization: \\ The electrokinetic response of a porous particle to an external electric field}
\end{center}

\setcounter{equation}{0}
\setcounter{figure}{0}
\setcounter{section}{0}
\setcounter{table}{0}
\setcounter{page}{1}

\renewcommand{\thesection}{S.\arabic{section}}
\renewcommand{\thesubsection}{\thesection.\arabic{subsection}}
\makeatletter 
\def\tagform@#1{\maketag@@@{(S\ignorespaces#1\unskip\@@italiccorr)}}
\makeatother
\makeatletter
\makeatletter \renewcommand{\fnum@figure}
{\figurename~S\thefigure}
\makeatother
\makeatletter \renewcommand{\fnum@table}
{\tablename~S\thetable}
\makeatother
%\makeatletter \renewcommand{\fnum@equation}
%{\equationname~S\theequation}
%\makeatother

\section{Derivation of governing equations using modified Donnan and Gouy-Chapman models}

Ion transport in a porous media with bimodal pore size distribution, filled with a symmetric and binary electrolyte $z_{+}=z_{-} \equiv z>0$, under diffusion,a advection and electro-migration, can be described by the following Nernst-Planck equations \cite{biesheuvel2012electrochemistry}
\begin{equation}
\begin{split}
	\dfrac{\partial (p_{m}^{}c_{m}^{\pm}+p_{M}^{}c_{M}^{\pm})}{\partial t}+ \vec{\nabla} \cdot \left( -p_{M}\left( D_{M}^{\pm} \vec{\nabla}c_{M}^{\pm} \pm F z_{}^{\pm} b_{M}^{\pm} c_{M}^{\pm} \vec{\nabla} \varphi_{M}^{}\right)  \right)=0,
\label{NPPorous}
\end{split}
\end{equation}
where $p_{m}$, $p_{M}$ denote the porosities in the micro- (m) and macro-  (M) pores regions, respectively, 
$D$ is the diffusion coefficient, $b$ is the effective electrophoretic mobility. Neglecting transport processes taking place in the micropore region,
Eq.(S\ref{NPPorous}) can be expressed as transport equations for the macropore region, with source terms representing ions electrosorption and expulsion from the micropores,
\begin{subequations}
\begin{align}
	\frac{\partial c_{M}^{\pm}}{\partial t}+ \frac{1}{p_{M}^{}}\vec{\nabla} \cdot \left( p_{M} \vec{J}_{M}^{\pm}  \right)=-\frac{p_{m}}{p_{M}}\frac{\partial c_{m}^{\pm}}{\partial t}
\\
	\vec{J}_{M}^{\pm} = - D_{M}^{\pm} \vec{\nabla}c_{M}^{\pm} \mp F z_{}^{\pm} b_{M}^{\pm} c_{M}^{\pm} \vec{\nabla} \varphi_{M}^{ }.
\label{NPDonnan}
\end{align}
\end{subequations}
In the bulk, a similar equation holds, except with $p_{M}=1$, vanishing source terms and advection field.
% Taking half sum and half difference, we are led to
% \begin{subequations}
% \begin{align}
% 	\dfrac{\partial c_{M}}{\partial t}-\nabla^{2}c_{M}=-\https://preview.overleaf.com/public/ydprmyyfnqgv/images/83a466c81e3008a76951528679c2859b773bb484.jpegdfrac{\partial c_{m}}{\partial t}
% \\https://preview.overleaf.com/public/ydprmyyfnqgv/images/83a466c81e3008a76951528679c2859b773bb484.jpeg
% 	\vec{\nabla} \cdot \left( b_{M} c_{M} \vec{\nabla} \varphi_{M} \right)=\dfrac{\partial \rho_{m}}{\partial t},
% \end{align}
% \label{NPhalfsumhalfdifference}
% \end{subequations}
% where similar equations hold in the bulk without the source/sink terms on the right hand side.
Utilizing Einstein's relation in the macropore region and in the bulk (B)
\begin{equation}
	b_{M,B}^{\pm}=\dfrac{D_{M,B}^{\pm}}{R T},
\end{equation} 
allows to rewrite the current density as an electrochemical potential gradient, implicitly defined by 
\begin{equation}
	\vec{J}_{M,B}^{\pm}= -\dfrac{D_{M,B}^{\pm}}{R T}  c_{M,B}^{\pm} \vec{\nabla} \mu_{M,B}^{\pm}+ c_{M,B}^{\pm} \vec{u},
\end{equation}
where $\vec{u}$ is advection field in the bulk which is assumed to satisfy no-penetration condition into the porous region.
Assuming chemical equilibrium between macropores and micropores for both species, which implies equality of chemical potentials in the two regions,
\begin{equation}
	\mu_{m}^{(0),\pm}+R T \ln \left( c^{\pm}_{m}\right) \pm F z \varphi_{m}=\mu^{(0),\pm}_{M}+R T \ln(c_{M}^{\pm})\pm F z \varphi_{M},
\label{ChemicalPotentiaEqui}
\end{equation}
and furthermore assuming for simplicity that the excess constant potentials for both species are equal, $\Delta \mu \equiv \mu_{m}^{(0),+}-\mu^{(0),+}_{M}=\mu_{m}^{(0),-}-\mu^{(0),-}_{M}$, we obtain the following relation between the excess of electrostatic potential (Donnan potential $\Delta \varphi_{d}$) and the ionic species concentrations in the micropores region
\begin{equation}
	\Delta \varphi_{d} \equiv \varphi_{m}-\varphi_{M} =\dfrac{V_{T}}{2 z}\ln \left( \dfrac{c_{m}^{-}}{c_{m}^{+}} \right),
\label{Donnan}
\end{equation} 
where $V_{T}=RT/F$ is the thermal voltage, and the constants $R$ and $F$ stand for the universal gas constant and Faraday's constant, respectively. 
Utilizing Eq.(S\ref{ChemicalPotentiaEqui}) and Eq.(S\ref{Donnan}) then leads to the following relation between species concentrations in the macropores and micropores
\begin{equation}
	c_{m}^{\pm}=c_{M}^{\pm} e^{-\dfrac{\Delta \mu/F \pm z \Delta \varphi_{d}}{V_{T}}}.
\label{DonnanBoltzmann}
\end{equation}
Utilizing Eq.(S\ref{DonnanBoltzmann}) we then obtain the following relations for the half charge density $\rho_{m,M} = F z (c_{m,M}^{+}-c_{m,M}^{-})/2$ and half neutral salt concentration $c_{m,M} = (c_{m,M}^{+}+c_{m,M}^{-})/2$
\begin{subequations}
\begin{align}
	c_{m}&=c_{M} e^{-\frac{\Delta \mu}{F V_{T}}} \cosh \left( \dfrac{z \Delta \varphi_{d}}{V_{T}} \right)
\\
	\rho_{m}&=-Fz c_{M} e^{-\frac{\Delta \mu}{F V_{T}}} \sinh \left( \dfrac{z \Delta \varphi_{d}}{V_{T}} \right),
\end{align}
\label{ChargeSaltDonnan}
\end{subequations}
as well as the relation
\begin{equation}
	c_{m}^{2}-\rho_{m}^{2}/(F z)^{2}= \left( c e^{-\frac{\Delta \mu}{F V_{T}}} \right)^{2}.
\end{equation}  
Substituting Eq.(S\ref{DonnanBoltzmann}) into Eq.(S\ref{NPDonnan}), leads to 
\begin{equation}
	\dfrac{\partial c_{M}^{\pm}}{\partial t}+ \dfrac{1}{p_{M}^{}}\vec{\nabla} \cdot \left( -p_{M}\left( D_{M}^{\pm} \vec{\nabla}c_{M}^{\pm} \mp F z_{}^{\pm} b_{M}^{\pm} c_{M}^{\pm} \vec{\nabla} \varphi_{M}^{} \right)  \right)=
	-\dfrac{p_{m}}{p_{M}} e^{-\frac{\Delta \mu}{e V_{T}}} \dfrac{ \partial \left( c_{M}^{\pm} e^{\mp \frac{ z \Delta \varphi_{d}}{V_{T}}}\right)}{\partial t},
\label{NPDonnanExplicit}
\end{equation}
which under the assumption of electroneutraility within the macropores region,  $\rho_{M}=0$, can be written for neutral salt concentration and charge density as
%\begin{equation}
%\begin{split}
%	&\dfrac{\partial c_{M}^{}}{\partial t}+ \dfrac{1}{p_{M}^{}}\vec{\nabla} \cdot \left( -p_{M}\left( D_{M}^{} \vec{\nabla}c_{M}^{} + b_{M}^{} \rho_{M}^{} \vec{\nabla} \varphi_{M}^{} \right)  \right)=
%\\
%	&-\dfrac{p_{m}}{p_{M}} e^{-\frac{\Delta \mu}{e V_{T}}} \dfrac{ \partial \left( c_{M}^{} \cosh \left( \frac{ z \Delta \varphi_{d}}{V_{T}} \right) \right)}{\partial t}
%\\
%	&\dfrac{\partial \rho_{M}^{}}{\partial t}+\dfrac{1}{p_{M}^{}} \vec{\nabla} \cdot \left( -p_{M}\left(  D_{M}^{} \vec{\nabla}\rho_{M}^{} +e^{2} z^{2} b_{M}^{} c_{M}^{} \vec{\nabla} \varphi_{M}^{} \right)  \right)=
%\\
%	&-\dfrac{p_{m}^{}}{p_{M}^{}}e^{-\frac{\Delta \mu}{e V_{T}}} \dfrac{\partial \left( c_{M}^{} \sinh \left( \frac{ z \Delta \varphi_{d}}{V_{T}} \right) \right) }{\partial t}.
%\end{split}
%\end{equation} 
\begin{equation}
\begin{split}
	\dfrac{\partial c_{M}^{}}{\partial t}- \dfrac{1}{p_{M}^{}}\vec{\nabla} \cdot \left( p_{M} D_{M}^{} \vec{\nabla}c_{M}^{} \right)=
-\dfrac{p_{m}}{p_{M}} e^{-\frac{\Delta \mu}{F V_{T}}} \dfrac{ \partial \left( c_{M}^{} \cosh \left( \frac{ z \Delta \varphi_{d}}{V_{T}} \right) \right)}{\partial t}
\\
	\dfrac{1}{p_{M}^{}} \vec{\nabla} \cdot \left( -p_{M} F^{2} z^{2} b_{M}^{} c_{M}^{} \vec{\nabla} \varphi_{M}^{}  \right)=
-\dfrac{p_{m}^{}}{p_{M}^{}}F z e^{-\frac{\Delta \mu}{F V_{T}}} \dfrac{\partial \left( c_{M}^{} \sinh \left( \frac{ z \Delta \varphi_{d}}{V_{T}} \right) \right) }{\partial t}.
\label{NPHalfSumHalfDifference}
\end{split}
\end{equation} 
Finally, we relate the Donnan potential to surface properties by expressing the potential drop between the solid electrode, $\varphi_{e}$, and the potential in the macropore solution, $\varphi_{M}$, via
\begin{equation} 
	\varphi_{e}-\varphi_{M}=\left( \varphi_{e}-\varphi_{m} \right)-(\varphi_{M}-\varphi_{m})=\Delta \varphi_{s}+\Delta \varphi_{d},
\label{PotentialDrop}
\end{equation}
where $\Delta \varphi_{s}$ is the so called Stern potential which represents the potential difference between the electrode surface and the micropore solution. Denoting $C_{s}$ as the volumetric micropore Stern layer capacitance, and assuming a linear capacitor relation between the depth averaged charge density, $\rho_{m}$, and the Stern potential 
\begin{equation}
	\rho_{m}=-\dfrac{1}{2} C_{s} \Delta \varphi_{s}
\label{SternCapacity}
\end{equation} 
and then substituting Eq.(S\ref{ChargeSaltDonnan}b) and Eq.(S\ref{SternCapacity}) into
Eq.(S\ref{PotentialDrop}), we obtain
\begin{equation}
	\varphi_{e}-\varphi_{M}=\Delta \varphi_{d}+ \dfrac{2 F z}{C_{s}}  e^{-\frac{\Delta \mu}{e V_{T}}} c_{M} \sinh \left( \dfrac{z \Delta \varphi_{d}}{V_{T}} \right).
\label{DonnanRelatesToOther}
\end{equation}

On the boundary between the porous matrix and the bulk, $\Gamma$, ionic species concentrations and the electric potential are continuous 
\begin{equation}
	c_{M}^{\pm} \Big \vert_{\Gamma^{-}}=c_{B}^{\pm} \Big \vert_{\Gamma^{+}}; \quad \varphi_{M}^{\pm} \Big \vert_{\Gamma^{-}}=\varphi_{B}^{\pm} \Big \vert_{\Gamma^{+}},
\label{BCPotentials}
\end{equation}
while matter and charge conservation imply equality of the fluxes in the normal ($n$) and tangential ($\theta$) directions
\begin{equation}
\begin{split}
	p_{M} \dfrac{\partial \varphi_{M}}{\partial n} \Big \vert_{\Gamma^{-}}= \dfrac{\partial \varphi_{B}}{\partial n} \Big \vert_{\Gamma^{+}}; \quad  p_{M} D_{M}^{\pm} \dfrac{\partial c^{\pm}_{M}}{\partial n} \Big \vert_{\Gamma^{-}}=D_{B}^{\pm}\dfrac{\partial c^{\pm}_{B}}{\partial n} \Big \vert_{\Gamma^{+}}
    \\
    p_{M} \dfrac{\partial \varphi_{M}}{\partial \theta} \Big \vert_{\Gamma^{-}}= \dfrac{\partial \varphi_{B}}{\partial \theta} \Big \vert_{\Gamma^{+}}; \quad  p_{M} D_{M}^{\pm} \dfrac{\partial c^{\pm}_{M}}{\partial \theta} \Big \vert_{\Gamma^{-}}=D_{B}^{\pm}\dfrac{\partial c^{\pm}_{B}}{\partial \theta} \Big \vert_{\Gamma^{+}},
\end{split}    
\label{BCDerivativesSurface}
\end{equation}  
where $\Gamma^{\pm}$ denote the inner $(-)$ and outer $(+)$ sides of the boundary.

A similar model for capacitive charging can be obtained using the Gouy-Chapman capacitive model \cite{hunter2001foundations2}, which describes transport in a porous material with thin, non-overlapping, EDLs. It requires the consideration of only one type of pores with porosity $p$, defined as the ratio of void volume, $V_{p}$, to the total volume (void and solid), $V$, via $p=V_{p}/V$. The transport equations analogous to Eq.(S\ref{NPHalfSumHalfDifference}) are then given by
\begin{equation}
\begin{split}
	\dfrac{\partial c_{M}^{}}{\partial t}- \dfrac{1}{p}\vec{\nabla} \cdot \left( p D_{M}^{} \vec{\nabla}c_{M}^{} \right)=
-\dfrac{a_{p}}{p} \dfrac{1}{F z} \dfrac{ \partial \left( \sqrt{8 \varepsilon R T c_{M}^{ }}\sinh^{2} \left( \frac{ z (\varphi-\varphi_{s}) }{4 V_{T}} \right) \right)}{\partial t}
\\
	\dfrac{1}{p} \vec{\nabla} \cdot \left( -p F^{2} z^{2} b_{M}^{} c_{M}^{} \vec{\nabla} \varphi_{M}^{}  \right)=
- \dfrac{a_{p}}{p} \dfrac{\partial \left( \sqrt{8 \varepsilon R T c_{M}^{ }} \sinh \left( \frac{ z (\varphi-\varphi_{s})}{2 V_{T}} \right) \right) }{\partial t},
\label{NPHalfSumHalfDifferenceGC}
\end{split}
\end{equation}
where $a_{p}=A_{p}/V$ is the interfacial area of the pores area, $A_{p}$, per unit volume, $V$, and the subscript $M$ explicitly indicates that the equations are written within the pore region. Similar equations without the source/sink hold in the bulk, and the matching conditions Eq.(S\ref{BCDerivativesSurface}) hold for this model as well.

\section{Scaling and non-dimensional equations}

% It is useful to define the following quantites. Porosity, $p$, defined as  the ratio of void volume, $V_{p}$, to the total volume (void and solid), $V$, via $p=V_{p}/V$; pore thickness, $h_{p}$, defined as the ratio of the pore volume to the pore surface area, $A_{p}$, via $h_{p}=V_{p}/A_{p}$. These are related via $h_{p}=\left(V_{p}/V\right) \cdot \left( V/A_{p} \right)=p/a_{p}$, where 
% \begin{equation}
% 	a_{p}=A_{p}/V
% \label{PoresArea}
% \end{equation}
% has dimension of $\text{length}^{-1}$ and stands as the surface (interfacial) area of the pore walls per unit volume.
% $2 D F z c_{0}/a$ is the current scale; 

% $2 D F^{2}z^{2} c_{0}/R T$ is the conductivity scale;

% $\epsilon_{p} = a_{p} \lambda_{D}= p \lambda_{D}/h_{p}$ is the volume fraction occupied by the EDL; 

% $p$ is the average porosity; 

% $h_{p}=p/a_{p}$ is the average pore
% size; 

% $\lambda_{D}=\sqrt{\epsilon R T/ (2 F^{2} z^{2} c_{0})}$
% is the Debye screening length; 

% $V_{T}/(z\lambda_{D})=\sqrt{2 \epsilon R T c}$ is the electric charge scale. In Angstroms Debye length is given by $\lambda_{D}=\dfrac{3.05 \AA}{\sqrt{c_{0}[M]}}$ 

Using the scaling
\begin{equation}
	c \rightarrow c_{0} c, \quad \vec{r} \rightarrow a \vec{r}, \quad \vec{\nabla} \rightarrow \dfrac{1}{a}\vec{\nabla},  \quad t \rightarrow\dfrac{a^{2}}{D} t, \quad \varphi \rightarrow \dfrac{V_{T}}{z} \varphi, \quad b \rightarrow \dfrac{D}{R T} b, \quad \Delta \mu \rightarrow V_{T} F \Delta \mu
\end{equation}
where $c_{0}$ is the initial concentration , $a$ is the disk radius, and $V_{T}$ is the thermal voltage,
% where $\lambda_{D}/a$ factor in the scaling of $I$ represents the sensible correction needed to obtain the correct $RC$ time scale \cite{bazant2004diffuse} relevant for charging, 
we recast the governing equations of the mD model, Eq.(S\ref{NPHalfSumHalfDifference}) and Eq.(S\ref{DonnanRelatesToOther}), in a dimensionless form as
\begin{equation}
\begin{split}
	\dfrac{\partial c_{M}^{}}{\partial t}-  \nabla^{2} c_{M}^{} =
-\dfrac{p_{m}}{p_{M}} e^{-\Delta \mu } \dfrac{ \partial \left( c_{M}^{} \cosh \left( \Delta \varphi_{d} \right) \right)}{\partial t}
\\
	 \vec{\nabla} \cdot \left(b_{M}^{} c_{M}^{} \vec{\nabla} \varphi_{M}^{}  \right)=
\dfrac{p_{m}^{}}{p_{M}^{}}e^{-\Delta \mu } \dfrac{\partial \left( c_{M}^{} \sinh \left( \Delta \varphi_{d} \right) \right) }{\partial t},
\label{NPHalfSumHalfDifferenceNonDim}
\end{split}
\end{equation}
and
\begin{equation}
	\varphi_{e}-\varphi_{M}=\Delta \varphi_{d}+ \dfrac{2 F c_{0} z^{2}}{V_{T} C_{s}} e^{-\Delta \mu} c_{M}  \sinh \left( \Delta \varphi_{d} \right),
\label{DonnanRelatesToOtherNonDim}
\end{equation}
respectively.

For the GC model, the governing non-dimensional transport equations take the form
\begin{equation}
\begin{split}
	\dfrac{\partial c_{M}^{}}{\partial t}- \nabla^{2} c_{M}^{}=
-4 \varepsilon_{p} \dfrac{ \partial \left( \sqrt{c_{M}^{ }}\sinh^{2} \left( \frac{ \varphi-\varphi_{s} }{4} \right) \right)}{\partial t}
\\
	\vec{\nabla} \cdot \left( b_{M}^{} c_{M}^{} \vec{\nabla} \varphi_{M}^{}  \right)=
-2 \varepsilon_{p} \dfrac{\partial \left(\sqrt{c_{M}^{ }} \sinh \left( \frac{ \varphi-\varphi_{s}}{2} \right) \right) }{\partial t},
\label{NPHalfSumHalfDifferenceGCnonDim}
\end{split}
\end{equation}
where we have defined $\varepsilon_{p}=\lambda_{D}/h_{p}$.
Here, we used the expression for the Debye length, $\lambda_{D}=\sqrt{\varepsilon R T/(2 F^{2} z^{2} c_{0})}$, for symmetric and binary electrolytes, and the 
pore thickness, $h_{p}$, defined as the ratio of the pore volume to the pore surface area, $A_{p}$, via $h_{p}=V_{p}/A_{p}$. These are related via $h_{p}=\left(V_{p}/V\right) \cdot \left( V/A_{p} \right)=p/a_{p}$. 
The dimensionless expressions for $c_{m}$ and $\rho_{m}$ in mD and GC models, respectively, are given by
\begin{equation}
\begin{split}
	c_{m}= c_{M}e^{-\Delta \mu}\cosh \left( \Delta \varphi_{d} \right)
\\
	\rho_{m}= c_{M}e^{-\Delta \mu}\sinh \left( \Delta \varphi_{d} \right).
\end{split}
\label{DN}
\end{equation}
and
\begin{equation}
\begin{split}
	c_{m}=4 \sqrt{c_{M}} \sinh^{2} \left( \dfrac{\varphi-\varphi_{s}}{4} \right)
\\
	\rho_{m}=-2 \sqrt{c_{M}} \sinh \left( \dfrac{  \varphi-\varphi_{s}}{2} \right).
\end{split}
\label{GC}
\end{equation}

% \begin{equation}
% \begin{split}
% 	\dfrac{\partial c}{\partial t}&+\vec{\nabla} \cdot \left( -\vec{\nabla}c + \rho \vec{\nabla} \varphi  \right)= -\varepsilon_{p}\dfrac{\partial w}{\partial t}
% \\
% 	\dfrac{\partial \rho}{\partial t}&+\vec{\nabla} \cdot \left( - \vec{\nabla}\rho +  c \vec{\nabla} \varphi  \right)=-\varepsilon_{p}\dfrac{\partial q}{\partial t},
% \end{split}
% \label{NPpairNonDimensionalPorosive}
% \end{equation}
% where $\varepsilon_{p}=\lambda_{D}/h_{p}$, and $w$, $q$ are described by a relevant capacitive model.
% Note the non-homogeneous sink terms on the right hand side of Eq.(\ref{NPpairNonDimensionalPorosive}) preclude transport effects within the EDL, which are taken into account in a 1D straight pores model in \cite{mirzadeh2014enhanced}.

\section{Expansion of the governing equations for early times}

In this section we obtain expressions for the
coefficients $A$ and $B$ introduced in Eq.(4a-b) which describe capacitive charging for both Gouy-Chapman and modified Donnan capacitive models. 

Consider the (non-dimensional) transport equations for the neutral salt concentration and charge density in the macropore region, given by 
Eq.(S\ref{NPHalfSumHalfDifferenceNonDim}), where specific expressions for $c_{m}$ and $\rho_{m}$ for Gouy-Chapman and modified Donnan capacitive models are given, respectively, by Eq.(S\ref{DN}) and Eq.(S\ref{GC}).
In the modified Donnan model we must also consider the additional relation, Eq.(S\ref{DonnanRelatesToOtherNonDim}), between the sums of the Donnan and Stern potentials and the total potential difference between the surface of the porous solid and the macropore region.
We perform a Taylor expansion of the neutral salt concentration and electric potential for small times (after $t=0$), $c=1+ \delta c + \delta^{2}c + \ldots$ and $\varphi-\varphi_{e}=\delta \varphi + \delta \varphi^{2} + \ldots$, where $\delta$ stands for infinitesimal change due to change $t \rightarrow t+\delta t$, and obtain relations for each order.

\textbf{Modified Donnan model} 

Expanding the right hand side of Eq.(S\ref{NPHalfSumHalfDifferenceNonDim}a) and Eq.(S\ref{NPHalfSumHalfDifferenceNonDim}b), we obtain
the following relations between $\delta \varphi$ and $\delta^{2}c$
\begin{subequations}
\begin{align}
	\text{1st order: \quad}& \dfrac{\partial \delta c}{\partial t}-\nabla^{2} \delta c=0
\\
	\text{2nd order: \quad}& \dfrac{\partial^{2} \delta^{2} c}{\partial t^{2}}-\nabla^{2} \delta^{2} c=
    \gamma^{2} e^{-\Delta \mu}\dfrac{p_{m}}{p_{M}}\delta \varphi \dfrac{\partial \delta \varphi}{\partial t},
\end{align}
\label{cexpansionD}
\end{subequations}
and
\begin{equation}
\begin{split}
	\hspace{-0.75in}  \text{1st order: \quad}&  \nabla^{2} \delta \varphi = -\gamma e^{-\Delta \mu} \dfrac{p_{m}}{p_{M}}\dfrac{\partial \delta \varphi}{\partial t},
\end{split}
\label{phiexpansionD}
\end{equation}
respectively. 
Here, we have used the linearized relation Eq.(S\ref{DonnanRelatesToOtherNonDim}), given by $\varphi = \gamma \Delta \varphi_{D}$, where $\gamma$ is a constant explicitly given by
\begin{equation}
	\gamma=-\left( 1+ \beta \right), \quad \beta=\dfrac{2 F c_{0} z^{2}}{V_{T}C_{s}}e^{-\Delta \mu}
\label{Gamma}
\end{equation}
Here, we assumed $\delta \varphi_{D} \vert_{t=0}=0$ and set $\delta c=0$. The latter follows from 
Eq.(S\ref{cexpansionD}a) and $c\vert_{t=0}=c_{0}$.
Eq.(S\ref{cexpansionD}b) and Eq.(S\ref{phiexpansionD}) implicitly defines the coefficients $A,B$ as
\begin{equation}
	A=\gamma^{2}e^{-\Delta \mu}\dfrac{p_{m}}{p_{M}}, \quad B=\gamma e^{-\Delta \mu} \dfrac{p_{m}}{p_{M}}.
\end{equation}

\textbf{Gouy-Chapman model}

Expanding the right hand side of Eq.(S\ref{NPHalfSumHalfDifferenceGCnonDim}) leads to the following relations
\begin{equation}
\begin{split}
	\text{1st order: \quad}& \dfrac{\partial \delta c}{\partial t}-\nabla^{2} \delta c=0
\\
	\text{2nd order: \quad}& \dfrac{\partial \delta^{2} c}{\partial t}-\nabla^{2} \delta^{2} c= 
    \dfrac{\varepsilon_{p}}{2}\delta \varphi \dfrac{\partial \delta \varphi}{\partial t}.
\end{split}
\label{cexpansion}
\end{equation}
% Eq.(S\ref{cexpansion}a) implies 
% The first-order change in concentration $\delta c=\dfrac{\partial c}{\partial t} \delta t =\nabla^{2} \delta t$, which must vanish. 
Similarly expanding Eq.(S\ref{NPHalfSumHalfDifference}b), and setting $\delta c=0$, we obtain
\begin{equation}
\begin{split}
	 \text{1st order: \quad}& \nabla^{2} \delta \varphi = 
    -\dfrac{\varepsilon_{p}}{2}\dfrac{\partial \delta \varphi}{\partial t}.
\end{split}
\label{phiexpansion}
\end{equation}
which together with Eq.(S\ref{cexpansion}b) implicitly define the coefficients $A$ and $B$ as 
\begin{equation}
	A= \dfrac{\varepsilon_{p}}{2}, \quad B= -\dfrac{\varepsilon_{p}}{2}.
\end{equation}

%The latter can be estimated from the electric field equation for short times (see \cite{SuppInfo}), and is given by the transmission-line time scale $\tau_{TL}=p_{m}/p_{M} \cdot a^{2}/D \cdot e^{-\Delta \mu/F V_{T}}$ \cite{biesheuvel2010nonlinear, mirzadeh2014enhanced}. 
%\deleted{after a short transition time, $\tau_{TL}$, depletion of the bulk begins to limit the availability of the salt.
%The time scale for $\tau_{TL}$ can be obtained from scaling of the governing Eq.(\ref{NPWithDonnan}), (see \cite{SuppInfo} for derivation) and is given by $\tau_{TL}=p_{m}/p_{M} \cdot a^{2}/D \cdot e^{-\Delta \mu/F V_{T}}$,
%which is analogous to the transmission line time scale obtained from the GC model, $\lambda_{D}/h_{p} \cdot  l_{p}^{2}/D$,
% \cite{biesheuvel2010nonlinear, mirzadeh2014enhanced}.
%For $p_{m}/p_{M} \sim 1$, $\Delta \mu=0$, $a=1$ mm and $D=10^{-9}$ m$^{2}$/sec, $\tau_{TL}=1000$ sec.}

Scaling of equations (S\ref{phiexpansionD},S\ref{phiexpansion}) leads to the time scales $\lambda_{D}/h_{p} \cdot a^{2}/D$ for the  GC model and $p_{m}/p_{M} \cdot e^{-\Delta \mu/F V_{T}} \cdot a^{2}/D$ for the mD model. These describe the transmission-line time scale \cite{biesheuvel2010nonlinear, mirzadeh2014enhanced2}, which describes the rate of penetration of electric field into the particle, in the absence of changes in concentration.

\section{Comment about boundary conditions at the macropores/bulk interface}

Consider a cylindrical porous particle at the instant when the electron charges within the particle have already responded to the electric field, while the ions in the liquid still have not, and set this time as $t=0$. Under an applied potential difference $\pm V$ between the lines $x=\pm L/2$ the value of the electrostatic potential on the porous particle, $\varphi_{e}$, which is centered between the electrodes vanishes at $t=0$.
% \begin{equation}
% 	\varphi_{e}=\frac{1}{2}\left( \text{ } \int\limits_{\gamma_{-}} \vec{\nabla} \varphi \cdot d\vec{r} -\int\limits_{\gamma_{+}}
%     \vec{\nabla} \varphi \cdot d\vec{r} \right),
% \end{equation}
% where $\gamma_{+}$ ($\gamma_{-}$) is a path from the surface $x=L/2$ ($x=-L/2$) to the porous electrode. 
For a symmetric deionization process (i.e. equal salt fluxes at both poles, the value of the (electrically) floating electrostatic potential $\varphi_{e}$ remains zero for $t>0$,
\begin{equation}
	\varphi_{e}=0.
\label{FloatingValue}
\end{equation}
However, for a deionization process which is not left/right symmetric, the floating value of the potential at $t>0$ may change in time, even for a centered disk. The latter stems from left/right asymmetric changes in conductivity which lead to different potential drops from the porous particle to each of the electrodes. In the limit $a \ll L$ the relative changes in conductivity in the left and right regions are negligible. For simplicity we assume that Eq.(S\ref{FloatingValue}) holds at all times.

On the particle, we require surface concentration and electric potential continuity
\begin{equation}
	c_{M} \Big \vert_{R^{-}}=c_{B} \Big \vert_{R^{+}}; \quad \varphi_{M} \Big \vert_{R^{-}}=\varphi_{B} \Big \vert_{R^{+}},
\label{BCPotentials}
\end{equation}
as well as matter and charge conservation in the radial direction
\begin{equation}
	p_{M} \dfrac{\partial \varphi_{M}}{\partial r} \Big \vert_{R^{-}}= \dfrac{\partial \varphi_{B}}{\partial r} \Big \vert_{R^{+}}; \quad  \dfrac{p_{M} D_{M}}{D_{B}} \dfrac{\partial c_{M}}{\partial r} \Big \vert_{R^{-}}=\dfrac{\partial c_{B}}{\partial r} \Big \vert_{R^{+}},
\label{BCDerivativesCylinder}
\end{equation}  
and similar continuity of the fluxes along the tangential direction. 
The initial neutral salt concentration is constant in the entire domain
\begin{equation}
	c(\vec{r},0)=c_{0},
\end{equation}
and the initial electrostatic potential (in the bulk, or in the macropores), in dimensionless units, and in the limit of high conductivity of the porous solid, should be set as
\begin{equation}
\varphi(r,\theta) \vert_{t=0}=
   \begin{cases}
       \left( \dfrac{1}{r}-r\right) \cos(\theta) \quad &\text{for} \quad r>1
\\
       \hspace{0.42in} k r^{q}\cos(\theta) \quad &\text{for} \quad r\leq 1.
    \end{cases}
\label{BCElectricPotentialTime}
 \end{equation}
Here, $k$ and $q$ are two regulating parameters, which we artificially introduce in order to maintain consistency between the initial condition, Eq.(S\ref{BCElectricPotentialTime}), and the boundary conditions, Eq.(S\ref{BCDerivativesCylinder}).
The for the naive case of $k=0$ (the potential of the liquid within the macropores equals the potential of the solid) a discrepancy emerges, as can be seen by substituting the initial condition, Eq.(S\ref{BCElectricPotentialTime}), into Eq.(S\ref{BCDerivativesCylinder}), leading to a contradiction ($-2=0$). This discrepancy stems from the fact that the underlying Nernst-Planck equations for the bulk lack the component which drives ions to the EDL -  the details of the charging process are captured solely by the source/sink terms in the porous region.  In this respect the $k=0$ case is singular since it attempts to describes charging of the bulk/solid interface, solely due to $\vec{\nabla}\varphi$. One possible way to resolve the contradiction is to assume that the initial electrostatic potential already penetrates a short distance into the porous electrode, such that the EDL charging occurs due to the source/sink terms.
By setting $k =-2/q$ (with $q>2$) the corrected potential allows to maintain consistency at $r=R=1$, and at the same time introduces a negligible numerical error (by choosing a large enough $q$). Perhaps a more rigorous argument on the bulk/solid boundary conditions can be obtained by considering the bulk/solid interface as a region of finite width $\lambda_{p}$, which may eventually lead to a mixed (Robin) boundary conditions on the interface $\varphi \vert_{R^{+}}=\left(  \varphi+\lambda_{p} \partial \varphi/ \partial r \right) \vert_{R^{-}}$, and provide the missing derivative term.

\section{Details of experimental setup}

\begin{figure}[ht]
\centering
\includegraphics[scale=0.7]{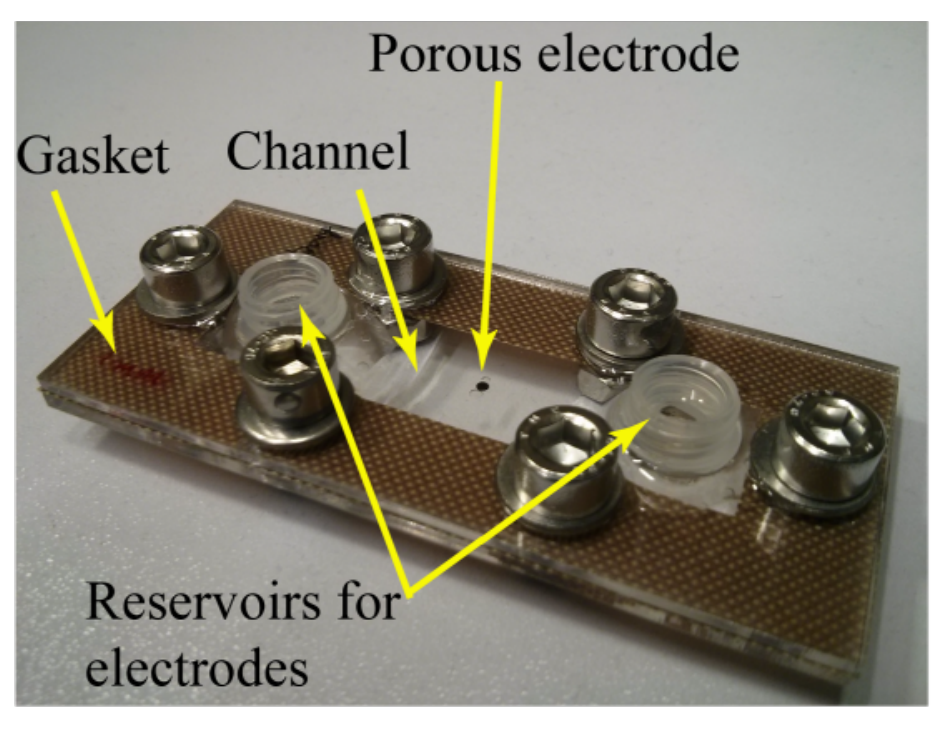}
         \caption{Photo of the device used for experimental study of ICCDI. The channel is formed by a $250$ $\mu$m thick gasket frame which is pressed between two $5$ mm thick transparent acrylic plates by six screws. The top acrylic plate has two reservoirs around two ciruclar apertures of $3$ mm diameter, for injection and removal of liquid into the channel. $1$ mm activated carbon, porous cylinder was placed at the center of the chamber. A typical experiment is performed by filling the channel with a liquid solution, and applying voltage through two eletrodes placed in each of the reservoirs}
\label{ExpPhoto}
\end{figure}

\textbf{Fabrication and materials:} The $15$ mm $\times$ $75$ mm $\times$ $250$ $\mu$m microfluidic chamber (see Fig.(S\ref{ExpPhoto})) was constructed from a $250$ $\mu$m thick  gasket frame (PTFE coated glass fiber, American Durafilm, Holliston, MA), pressed between two $5$ mm thick transparent acrylic plates (Yavin Plast, Haifa, Israel) of lateral dimension $40$ mm $\times$ $90$ mm) using six bolts.  $1$ mm activated carbon cylinders were cut from activated carbon sheets by using a biopsy puncher. The carbon sheets were custom made by Wetsus from activated carbon powder  (Axion Power International Inc., New Castle, PA) containing 85 wt porous carbon material, 10 wt polyvinylidene fluoride (PVDF), and 5 wt carbon black. The carbon cylinders were pressed between the two acrylic plates. 6 mm diameter through-holes were drilled into the top acrylic plate for fluidic access. On top of the holes, two plastic caps were glued (NOA68, Norland, Cranbury, NJ 08512) to serve as reservoirs.  

%%%%%%%%%%%%%%%%%%%       Fig.1      %%%%%%%%%%%%%%%%%

The electrolyte solution we used in the experiments is $100 $ $\mu$M sodium fluorescein (Sigma-Aldrich, St. Louis, MO) dissolved in deionized water (Millipore Milli-Q system, Billerica, MA).

\textbf{Equipment:} We applied voltage using a sourcemeter (model $2410$, Keithley Instruments, Cleveland, OH), connected to two platinum electrodes dipped into each of the reservoirs.  The fluroescence in the chamber was imaged using an inverted epifluorescent microscope (Ti-U, Nikon, Tokyo, Japan) equipped
with a metal halide light source (Intensilight, Nikon Japan) and a Chroma $49011$ filter-cube ($480/40$ nm excitation, $535/50$ nm emission and $510$ nm dichroic mirror). We used a $4\times$ objective ($\text{NA} = 0.13, \text{WD} = 17.2 \text{mm}$) for the experiments with a single activated carbon particle, and Nikon Plan Fluor objective $1 \times$ objective ($\text{NA} = 0.04, \text{WD} = 3.2 \text{mm}$, Plan UW, Nikon, Tokyo, Japan) for the experiments with the porous particles array.  Images were captured using a $14$ Bit, $1300
\times 1030$ pixel array CCD camera (Clara, Andor, Belfast, Ireland). We triggered the camera at intervals of $600$ ms with an exposure time of $200$ ms. We controlled the camera using NIS Elements software (v.4.11, Nikon) and processed the images with MATLAB (R2011b, Mathworks, Natick, MA).

\textbf{Experiment protocol:} Before each run, we washed the chamber thoroughly (approximately 5 min) with DI. We then filled the chamber with the sodium fluorescein solution and allowed $30$ min before initiating the electric field. Each carbon cylinder was used multiple times, and the chip was stored in DI when not in use.

% %%%%%%%%%%%%%%%%%%%       Fig.1      %%%%%%%%%%%%%%%%%
% \begin{figure}[ht!]
% \centering
% \includegraphics[scale=0.7]{ExpSetupICCDI2_artboard2-02.eps}
%           \caption{Photo of the device used for experimental study of ICCDI. The channel is formed by a 250 [um] thick gasket frame which is pressed between two 5 [mm] thick transparent acrylic plates by six screws. The top acrylic plate has two reservoirs around two ciruclar apertures of 3 [mm] diameter, for injection and removal of liquid into the channel. 1 [mm] activated carbon, porous cylinder was placed at the center of the chamber. A typical experiment is performed by filling the channel with a liquid solution, and applying voltage through two eletrodes placed in each of the reservoirs}
% \label{ExpPhoto}
% \end{figure}

\section{Details of numerical simulation}

\subsection{Computational grid}

The computational domain is a square, with $(-L/2,-L/2)$, $(L/2,L/2)$ as two of its opposite vertices and $L=20$.  The grid is based on $100 \times 100$ rectangular elements with a non-uniform density, such that most of the elements are concentrated near the porousregion: both dimensions of the elements grow  using an arithmetic sequence from $0.03$ near the centerline to $0.3$ at the boundary.  We use second order (quadratic) elements, and solve the governing equations using the time-dependent solver with a time-step of $10^{-3}$.

\subsection{Governing equations and simulation parameters}

The full equations that govern the evolution of $c,\varphi,\Delta \varphi_{D}$ for the symmetric ($D^{+}=D^{-}=D$, $b^{+}=b^{-}=b$) and the non-symmetric (in the sense $D^{+}=D^{-}=D$, $b^{+} \neq b^{-}$) cases, respectively, are
\begin{equation}
\begin{split}
	\dfrac{\partial c}{\partial t}-\dfrac{1}{p_{M}}\vec{\nabla} \cdot \left(p_{M} f_{p} \right[ \vec{\nabla}c - (b^{+}-b^{-}) c \vec{\nabla} \varphi \left] \right) +& 
    r_{p} e^{-\Delta \mu} \left( \cosh(\Delta \varphi_{D}) \dfrac{\partial c}{\partial t} + c \sinh \left( \Delta \varphi_{D} \right) \dfrac{\partial \Delta \varphi_{D}}{\partial t}\right)=0
\\
	\dfrac{1}{p_{M}} \vec{\nabla} \cdot \left(p_{M} f_{p} c \vec{\nabla} \varphi \right)+r_{p} e^{-\Delta \mu}& \left(\sinh(\Delta \varphi_{D}) \dfrac{\partial c}{\partial t} + c \cosh(\Delta \varphi_{D}) \dfrac{\partial \Delta \varphi_{D}}{\partial t} \right) =0
\\
	\dfrac{\partial \varphi}{\partial t}+\dfrac{\partial \Delta \varphi_{D}}{\partial t}+ \beta & \left( \sinh(\Delta \varphi_{D}) \dfrac{\partial c}{\partial t}+ c \cosh(\Delta \varphi_{D}) \dfrac{\partial \Delta \varphi_{D}}{\partial t} \right) =0,
\end{split}
\label{GovEqSumulation}
\end{equation}
where $\beta$ is defined in Eq.(S\ref{Gamma}).
To avoid numerical issues associated with sharp interfaces defined smoothed step-like function to describe relevant quantities. 
Specifically, the  porosity functions, $p=p(x,y), p_{m}=p_{m}(x,y), p_{M}=p_{M}(x,y)$, are continuous through space, and defined using a smoothed step function, 
\begin{equation}
	H_{k}(x,y)=\frac{1}{2}\left( 1+\tanh\big[ k ( x^{2}+y^{2}-r_{0}^{2} ) \big] \right), 
\end{equation}
which interpolates between values unity and zero over a length scale $k^{-1}$, and in the limit $k \rightarrow \infty$ approaches the Heaviside step function.
In terms of $H_{k}$, the ratio of porosities, $r_{p}$, and the spatial variation in diffusivities, $f_{p}$, are given by  
\begin{equation}
\begin{split}
   r_{p}(x,y) = \dfrac{p_{m}(x,y)}{p_{M}(x,y)} = 
             \left( 1- \dfrac{p_{m}}{p_{M}} \right) H_{k}(x,y)+\dfrac{p_{m}}{p_{M}},
             \\
  f_{p}(x,y) = \left( D_{B}- D_{M} \right) H_{k}(x,y)+D_{M}
\end{split}
\end{equation}
% \begin{equation}
%                p = p(x,y) = (1/1.25) \cdot (H(x^{2}+y^{2}-r_{0})+0.25),
% \end{equation}
which smoothly interpolate between unity and $D_{B}$ outside the porous disk (of radius $r_{0}$), and $p_{m}/p_{M}$ and $D_{M}$ inside it, respectively.
The initial conditions are set to a uniform concentration through the entire domain, the electric potential outside the porous region is the sum of a dipole and a uniform electric field, and vanishes within the porous region. These are explicitly given by
\begin{equation}
\begin{split}
	c(t=0,\vec{r})=c_{0}, \quad \frac{\partial c}{\partial t} &\big \vert_{t=0}=0, 
\\
	\varphi(t=0,\vec{r})=\varphi_{0}(\vec{r}), \quad \frac{\partial \varphi}{\partial t} &\big \vert_{t=0}=0, 
\\
	\Delta \varphi_{d}(t=0,\vec{r})=0, \quad \frac{\partial \Delta \varphi_{d}}{\partial t} &\big \vert_{t=0}=0,
\end{split}
\label{InitCond}
\end{equation}
and
\begin{equation}
\begin{split}
	\varphi_{0}(x,y)=& \varphi_{i} \cdot \cos(\theta) \left( \dfrac{r^{2}_{0}}{r}-r \right) H_{k}(x,y)=
\\
          & \varphi_{i} \cdot \frac{x}{\sqrt{x^2+y^2}} \cdot \left(\frac{r_{0}^{2}}{\sqrt{x^2+y^2}}-\sqrt{x^2+y^2} \right) \cdot H_{k} \left(x,y \right).
\end{split}
\end{equation}
The salt concentration and the electrostatic potential satisfy Dirichlet boundary conditions on the lines $x= \pm L/2$ at all times, explicitly given by $c \vert_{x=\pm L/2} =c_{0}$, and $\varphi \vert_{x= \pm L/2} = \pm V$.

\textbf{Numerical values used}: 

The size of the square shaped domain: $L=20$

The radius of the porous region: $r_{0}=1$

The width of the smoothing region: $k^{-1}=10$ 

The electric potential of the porous disk at all times: $\phi_{e}=0$

Scale factor of the electric potential: $\varphi_{i}=-4$

The initial value of the concentration: $c_{0}=1$

Excess of chemical potential: $\Delta \mu=0$

Electrophoretic mobility: $b=1$ (for symmetric model) and $b^{+}-b^{-}=0.25$ (for asymmetric model)

Porosities: $p_{m}=0.3$, $p_{M}=0.4$

Ratio of diffusivities: $D_{M}/D_{B}=1/2$

Various constants: $\beta=10^{-4}$ which corresponds to Stern capacity $C_{s}=75$ MF/m$^{3}$, valence $z=1$, $F=96485.3$ C/mol and salt concentration $c_{0}=100$ $\mu$M.

\subsection{Simulation results}

\begin{figure}[ht]
\centering
\includegraphics[scale=0.3]{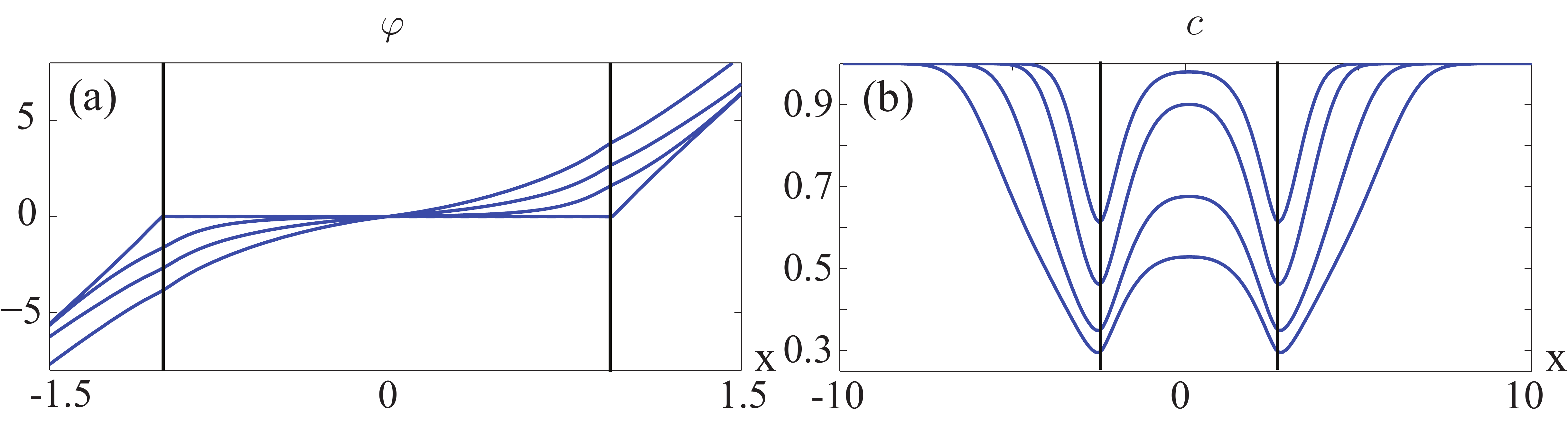}
         \caption{Simulation results of the symmetric model, presenting (a) electrostatic potential and (B) neutral salt concentration, at non-dimensional times $0$, $0.06$, $0.2$, $1$, and $0.1$, $0.2$, $0.5$, $1$, along a line which passes through the center of the disk parallel to the $x$-axis. For $x>0$ ($x<0$), and within the time interval we solved the numerical model, the electric potential is monotonically decreasing (increasing) function of time, whereas the concentration is an even decreasing function of time throughout the entire domain.}
\label{NumResults}
\end{figure}

Fig.(S\ref{NumResults}) presents simulation results for the electric potential and concentration along the line that passes through the center of the disk parallel to the x-axis. 
At early times the electrostatic potential is that of a dipole outside the porous region and vanishes inside, whereas the concentration is uniform in all regions. At later times the deionization invokes concentration changes near the porous-bulk interface (shown by lines $x=\pm 1$), which diffuse into both the inner and outer regions. At long times (relative to our normalization scale), the electrostatic potential approaches a uniform electric field and the charging process stops.
We find qualitative agreement against experimental data shown in Fig.(2b). 

%%%%%%%%%%%%%%%%%%%       Fig.1      %%%%%%%%%%%%%%%%%

%%%%%%%%%%%%%%%%%%%       Fig.2      %%%%%%%%%%%%%%%%%
\begin{figure}[ht]
\centering
\includegraphics[scale=0.5]{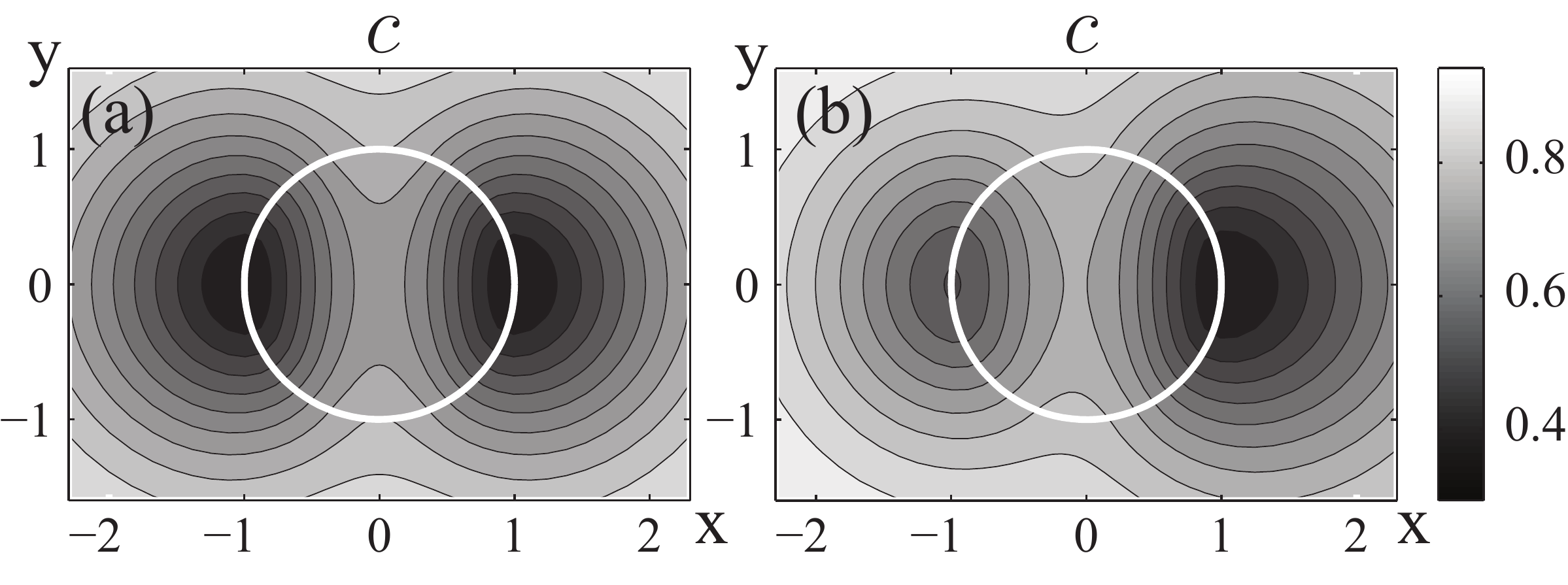}
         \caption{Numerical simulation results showing the concentration distribution inside and outside of the porous disk, at a non-dimensional time $0.5$ (scaled by $a^{2}/D$). (a) presents the symmetric case, while (b) shows the asymmetric case in which the two ions differ in their electrophoretic mobility.}
         
\label{NumResultsNoWhiteDisk}
\end{figure}

Fig.(S\ref{NumResultsNoWhiteDisk}) presents the salt distribution in the inner and outer regions for the symmetric (a) and asymmetric cases (b). It reveals that salt depletion within the porous is far more significant close to the poles than in the center. This is consistent with the fact that the potential difference between the liquid and the solid is smallest at the core of the particle. Fig.(S\ref{NumResultsStreamlines}) presents the salt distribution (colormap) and the electric field streamlines for a symmetric case at two intermediate times.  At early times, the electric field lines have a significant component perpendicular to the particle's outer surface ($t \ll \tau_{D}$). This normal component reduces at later times as the charging process proceeds.

%%%%%%%%%%%%%%%%%%%       Fig.3     %%%%%%%%%%%%%%%%%
\begin{figure}[ht!]
\centering
\includegraphics[scale=0.3]{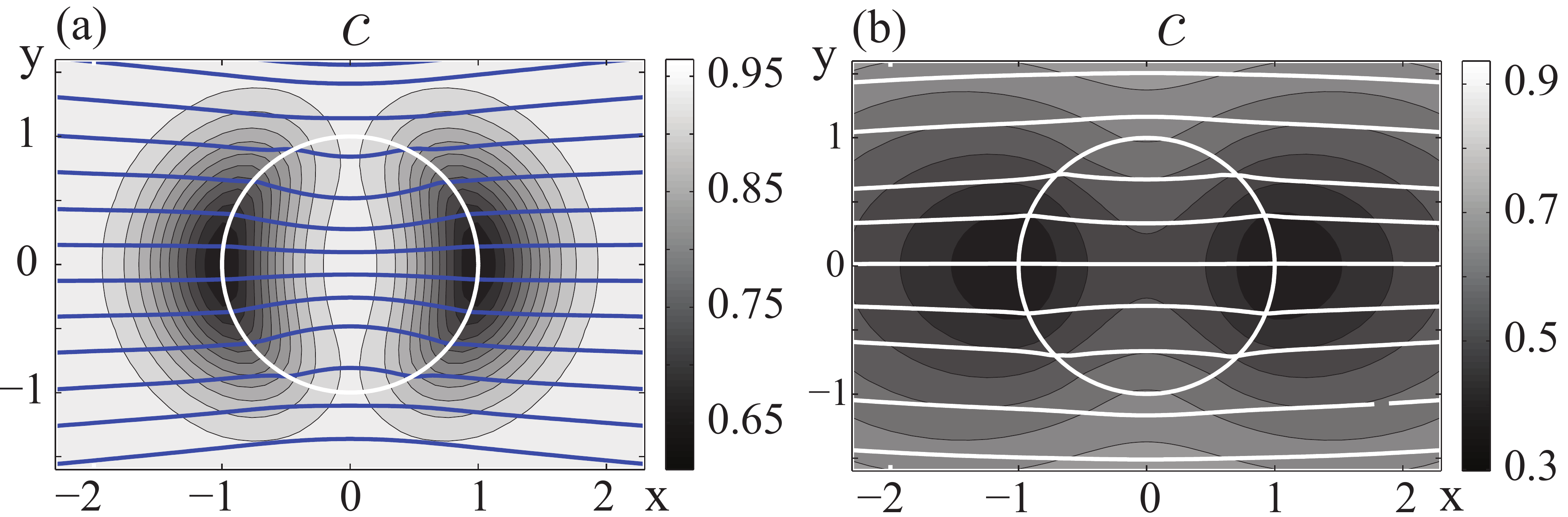}
         \caption{Numerical simulation result showing changes in the concentration distribution (colormap) and electric field lines for a symmetric electrolyte, at non-dimensional times (scaled by $a^{2}/D$) of (a) $0.2$ and (b) $1.5$.} 
\label{NumResultsStreamlines}
\end{figure}

%%%%%%%%%%%%%%%%%%%       Fig.4     %%%%%%%%%%%%%%%%%
% flux_circle_000_020_040_060_080_time_4_graph
\begin{figure}[ht!]
\centering
\includegraphics[scale=0.28]{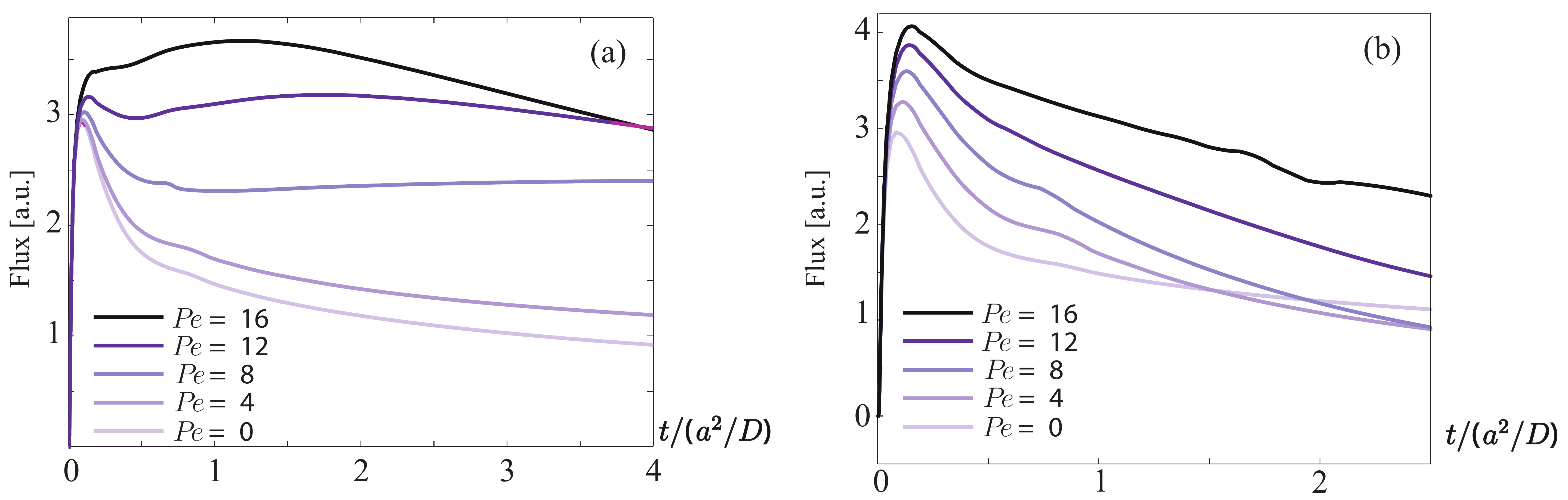}
         \caption{Numerical simulation results showing the total flux of salt into the porous region as a function of dimensionless time (scaled by $a^{2}/D$), for different Peclet ($Pe=a U/D$) numbers, where $U$ is the magnitude of maximal slip velocity on the disk boundary. (a) dipolar flow, (b) quadrupolar flow.}
\label{NumResultsStreamlines}
\end{figure}

\newpage

\section{Experimental Data}

\subsection{Multiple porous beads}

%  As conceptually demonstrated by our porous disk array experiment (Fig.(\ref{ICCDImatrix}d-f)), large volumes of liquid can be processed by multiple particles spaced approximately one diameter apart. In such a case, the time scales required to process the given volume does not differ much from the time required for a single particle to process a region of scale $a$. The processing time of the array is expected to be smaller by a factor of $a^{2}/L^{2}$ (ratio of diffusion times) compared to standard CDI setup having two electrodes at the edges of the volume.

\begin{figure}[ht]
\centering
\includegraphics[scale=0.40]{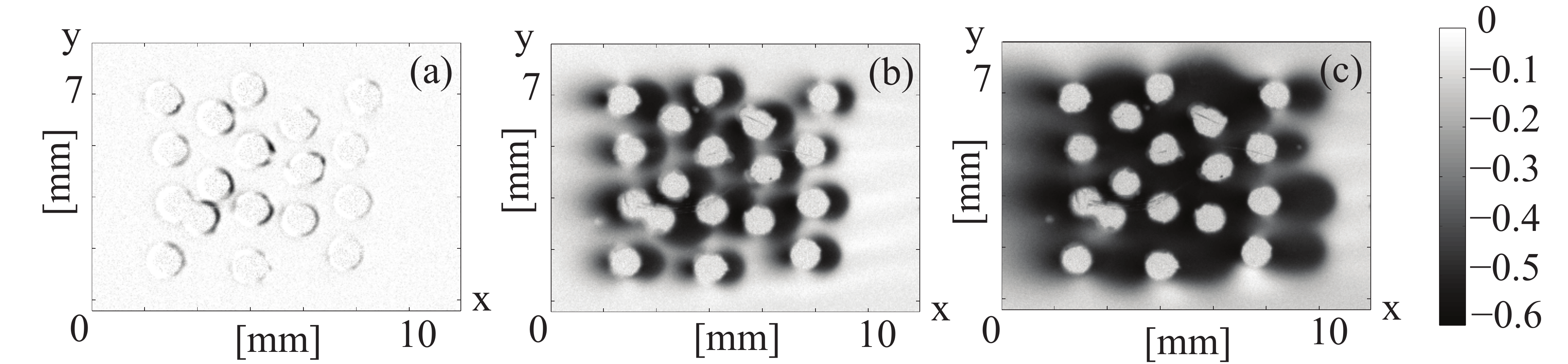}
         \caption{Experimental results demonstrating the use of ICCDI for depletion of salt in the bulk liquid by array of fixed, $1.2 \text{ }$mm diameter, porous carbon particles. Images show fluorescence signal of $100 \text{ } \mu$M sodium fluorescein under applied potential difference of $20 \text{ } V$ between right and left electrodes, at times (a) $6 s$, (b) $ 78 s$ and (c) $240 \text{ }$s. Each frame presents the change in fluorescence relative to the first frame, and is normalized by it for flat-field correction.} 
\label{Array}
\end{figure}

Fig.(S\ref{Array}a-c) presents a time-lapse experiment performed on an array of $18$ porous disks arranged in a staggered array, with a typical distance of $1$ mm between the disks. At $78$ s, clear interaction between the depletion fronts of the individual disks is observed, and by $240 \nobreak \text{ s}$ a continuous depletion region exists between the disks. i.e. a relatively large volume of the bulk can be processed using a set of porous electrodes, each sufficiently small to operate in an (induced) capacitive mode.
In such a case, the time scales required to process the given volume does not differ much from the time required for a single particle to process a region of scale $a$. The processing time of the array is expected to be smaller by a factor of $a^{2}/L^{2}$ (ratio of diffusion times) compared to standard CDI setup having two electrodes at the edges of the volume.

\subsection{Quasi-steady regime}

\begin{figure}[ht]
\centering
\includegraphics[scale=0.18]{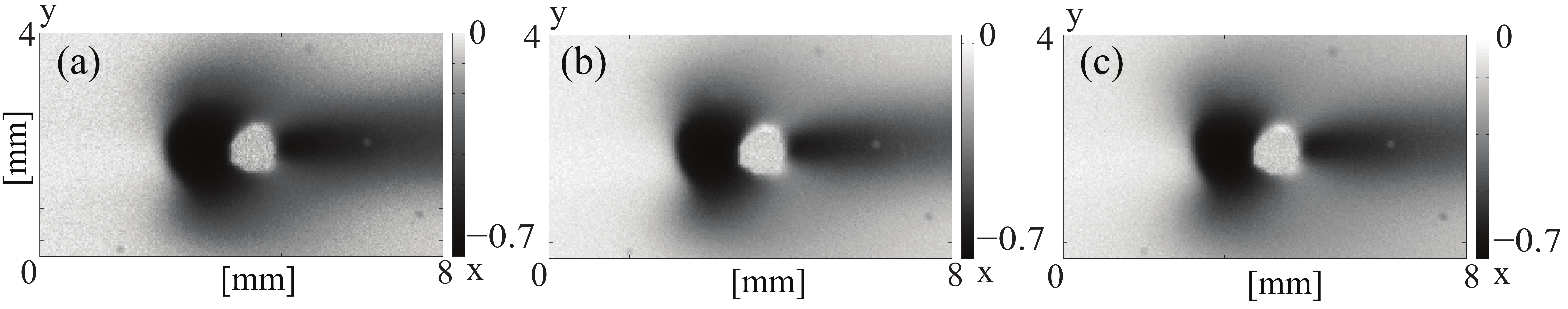}
         \caption{Raw fluorescence images showing the quasi-steady fluorescence distribution obtained from $100$ $\mu$M sodium fluorescein, around a $1.2$ mm diameter activated carbon disk after: (a) $720$ sec, (b) $920$ sec and (c) $1120$ sec. The potential difference is $20$ V along a distance of $75$ mm, and the corresponding uniform component of the electrical field has magnitude $260$ V/m and is oriented from left to right. We attribute the long wake seen on the right side of the particle to electro-osmotic flow present in our system.}
\label{160412_003_004}
\end{figure}

For sufficiently long times, the disk-shaped depletion region grows to a maximal size, at which the net flux of ionic species delivered by electromigration and diffusion is balanced by the charging rate of the micropores.
Fig.(\ref{160412_003_004}) presents a sequence of experimental images around a $1.2$ mm disk, taken every $3.3$ min after an initial transient of 12 min, showing practically identical fluorescene distributions around the porous particle.

%%%%%%%%%%%%%%%%%%%       Fig.4      %%%%%%%%%%%%%%%%%
Interestingly the left and right depletion regions have different characteristics. The depletion region around the negative pole on the left side, is disk-shaped and has a sharp transition between the dark to bright zones. The depletion region around a positive pole (on the right side), has a prolonged shape and a more diffused transition between the bright/dark zones.
We hypothesize that the described behavior stems from different charging rates of sodium fluorescein constituents (i.e. positively charged sodium ions and negatively charged Fluoroscein ions), as well as asymmetric pH distribution around the porous particle. 

\subsection{Discharge regime}

\begin{figure}[ht]
\centering
\includegraphics[scale=0.2]{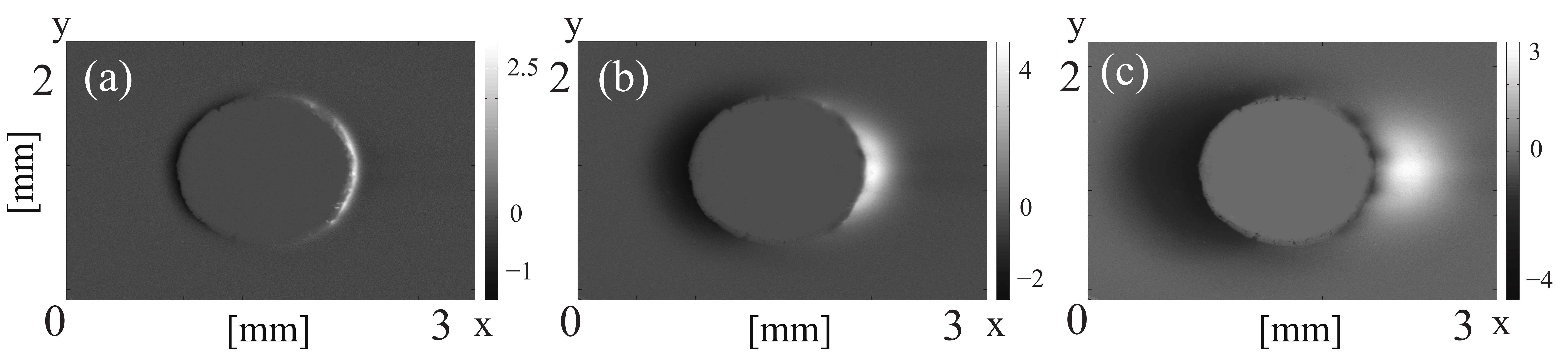}
         \caption{Raw fluorescence images showing the discharge process of $1.2$ mm diameter carbon disk previously charged with  $100$ $\mu$M sodium fluorescein for 5 min.  When the electric field is flipped, the initial discharge process results in release of fluorescein  from the right pole resulting in an observable intensity increase.  At later times, re-charging of the disk results in renewed depletion at the pole. The electrical field was $260$ V/m, oriented from right to left (a) 3 s, (b) 12 s, (c) 30 s.}
\label{FramesExp160131_002}
\end{figure}

After charging the porous particle for $5$ min, we turn off the electric field and replace the liquid in a chamber with a fresh sodium fluorescein solution, which again results in a uniform concentration distribution around the particle. We then apply an electric field in an opposite direction to the original charging field, and as expected observe enhanced fluorescence around the negative pole (i.e. positive pole during the preceding charging regime) resulting from the release of fluorescein into the liquid. Fig.(S\ref{FramesExp160131_002}) presents experimental images of the fluorescence at different times after the initiating discharge. After $30$ s, we again see a dark region forming at the right pole, due to its recharging (with sodium ions). 

%%%%%%%%%%%%%%%%%%%       Fig.5      %%%%%%%%%%%%%%%%%

\subsection{Terminology}

\begin{figure}[ht]
\centering
\includegraphics[scale=0.4]{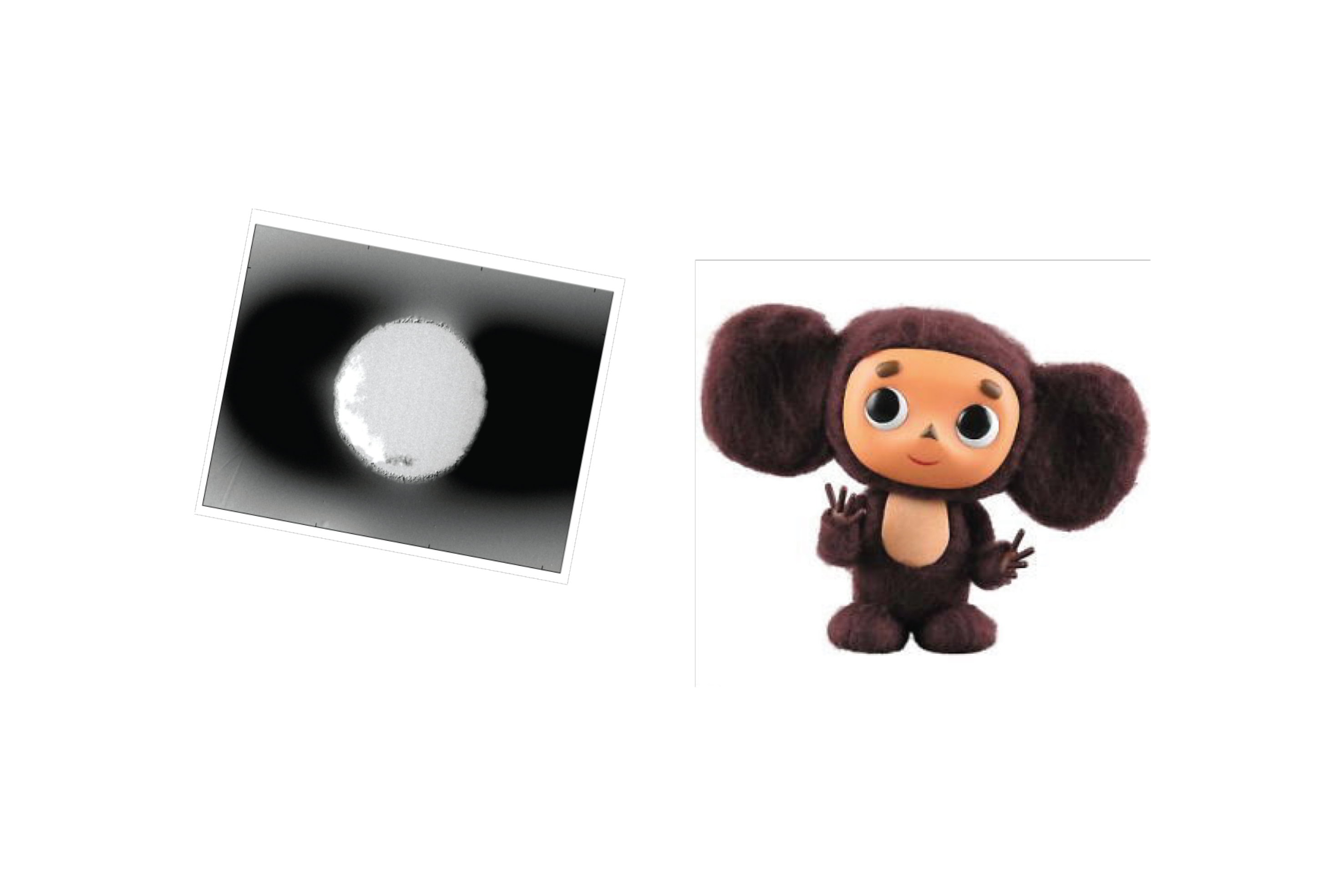}
         \caption{The shape of the depletion regions formed around a porous disk under ICCDI (left) share resemblance to Cheburashka, an iconic hero in Soviet literature and cartoons introduced by Soviet writer Eduard Uspensky on 1966.  The depletion regions can thus be conveniently referred to as "Cheburashka ears"\href{https://en.wikipedia.org/wiki/Cheburashka}{https://en.wikipedia.org/wiki/Cheburashka}}
\label{Cheburashka}
\end{figure}

We term the depletion region around the electrosorbing porous disk as `Cheburashka' ears. Our justification stems from the apparent similarity between the two, as shown in Fig.(S\ref{Cheburashka}).

%%%%%%%%%%%%%%%%%%%       Fig.5      %%%%%%%%%%%%%%%%%

% \subsection{ICCDI under pressure driven flow}

% The combined effect of electroadsorption and increasing of pressure driven advection, results in narrowing of the boundary layer which confines the low salt concentration region in a downstream wake and a region around the porous particle.
% Exact solution for large Peclet number and steady state adsorption was obtained via conformal methods techniques at \cite{bazant2003conformal}, and was numerically investigated for arbitrary Peclet number at

% %%%%%%%%%%%%%%%%%%%       Fig.6      %%%%%%%%%%%%%%%%%
% \begin{figure}[ht]
% \centering
% \includegraphics[scale=0.33]{160905_005_007_011.eps}
%          \caption{Raw fluorescence images showing the combined effect of ICCDI with pressure driven advection field from left to right, around $1.2$ mm diameter carbon disk at different flow rates; (a)  600 $\mu$m/sec cm/sec, (b) 1 cm/sec, (c) 4 cm/sec. The external electric field has magnitude of $260$ V/m and oriented from right to left. }
% \label{AdvectionFrames}
% \end{figure}

% \bibliography{ICCDI_SI}

%\clearpage
%

\end{document}